\documentclass[fleqn,usenatbib]{mnras}

\usepackage{newtxtext,newtxmath}
\usepackage[T1]{fontenc}
\usepackage{ae,aecompl}

\usepackage{graphicx}	% Including figure files
\usepackage{amsmath}	% Advanced maths commands
\newcommand{\HI}{H\,{\sc i}}

\newcommand{\Msun}{~M$_{\odot}$}

\newcommand{\kms}{~km\,s$^{-1}$}
\newcommand{\kkms}{km\,s$^{-1}$}

\newcommand{\vhel}{$v_{\rm hel}$}

\usepackage{xcolor}

\title[ASKAP reveals the filamentary radio tail structure of the Corkscrew Galaxy]{ASKAP reveals the radio tail structure of the Corkscrew Galaxy shaped by its passage through the Abell~3627 cluster}

\author[Koribalski et al. ]{B\"arbel S. Koribalski,$^{1,2}$\thanks{E-mail: Baerbel.Koribalski@csiro.au} Stefan W. Duchesne,$^{3}$ Emil Lenc,$^{1}$  Tiziana Venturi,$^{4,5}$  Andrea \newauthor Botteon,$^{4}$ Stanislav S. Shabala,$^{6}$ Tessa Vernstrom,$^{3,7}$ Ettore Carretti,$^{4}$ Ray P. Norris,$^{1,2}$ \newauthor Craig Anderson,$^{8}$ Andrew M. Hopkins,$^{9}$ C.J. Riseley,$^{10,4}$ Nikhel Gupta,$^{3}$ Velibor Velovi\'c$^{2}$ \\
$^{1}$Australia Telescope National Facility, CSIRO Astronomy and Space Science, P.O. Box 76, Epping, NSW 1710, Australia \\
$^{2}$School of Science, Western Sydney University, Locked Bag 1797, Penrith, NSW 2751, Australia \\
$^{3}$CSIRO Space and Astronomy, PO Box 1130, Bentley WA 6102, Australia \\
$^{4}$INAF-IRA, via P. Gobetti 101, I-40129, Bologna, Italy \\
$^{5}$Center for Radio Astronomy Techniques and Technologies, Rhodes University, Grahamstown 6140, South Africa \\
$^{6}$School of Natural Sciences, Private Bag 37, University of Tasmania, Hobart 7001, Australia \\
$^{7}$ICRAR, The University of Western Australia, 35 Stirling Hw, 6009 Crawley, Australia \\
$^{8}$Research School of Astronomy and Astrophysics, The Australian National University, ACT 2611, Australia \\
$^{9}$School of Mathematical and Physical Sciences, 12 Wally’s Walk, Macquarie University, NSW 2109, Australia  \\
$^{10}$Dipartimento di Fisica e Astronomia, Universit\`a degli Studi di Bologna, via P. Gobetti 93/2, 40129 Bologna, Italy
}

\date{Accepted XXX. Received YYY; in original form ZZZ}

\pubyear{2024}

\begin{document}
\label{firstpage}
\pagerange{\pageref{firstpage}--\pageref{lastpage}}
\maketitle

\begin{abstract}
Among the bent tail radio galaxies common in galaxy clusters are some with long, collimated tails (so-called head-tail galaxies) shaped by their interactions with the intracluster medium (ICM). Here we report the discovery of intricate filamentary structure in and beyond the $\sim$28\arcmin\ (570~kpc) long, helical radio tail of the Corkscrew Galaxy (1610--60.5, ESO\,137-G007), which resides in the X-ray bright cluster Abell~3627 ($D = 70$~Mpc). Deep radio continuum data were obtained with wide-field Phased Array Feeds on the Australian Square Kilometer Array Pathfinder (ASKAP) at 944~MHz and 1.4~GHz. The Corkscrew Galaxy is located 15\arcmin\ north of the prominent wide-angle tail (WAT) radio galaxy 1610--60.8 (ESO\,137-G006) near the cluster centre. While the bright (young) part of its radio tail is highly collimated, the faint (old) part shows increasing oscillation amplitudes, break-ups, and filaments. We find a stunning set of arc-shaped radio filaments beyond and mostly orthogonal to the collimated Corkscrew tail end, forming a partial bubble. This may be the first detection of a "proto-lobe" seen in 3D MHD simulations by Nolting et al. (2019), formed by the face-on impact of the Corkscrew Galaxy with a shock front in the cluster outskirts. Interactions of the radio galaxy tail with the ICM are likely responsible for the tail collimation and shear forces within the ICM for its increasingly filamentary structure. We also report the discovery of small ($\sim$20--30~kpc) ram-pressure stripped radio tails in four Abell~3627 cluster galaxies.
% 249 words

% The abstract should briefly describe the aims, methods, and main results of the paper. It should be a single paragraph not more than 250 words (200 words for Letters). No references should appear in the abstract.
\end{abstract}


%%%%%%%%%%%%%%%%%%%%%%%%%%%%%%%%%%%%%%%%%%%%%%%%%%

%%%%%%%%%%%%%%%%% BODY OF PAPER %%%%%%%%%%%%%%%%%%

\section{Introduction} 
\label{sec:intro}

\begin{figure*} % Figure 1
\centering
 \includegraphics[width=16cm]{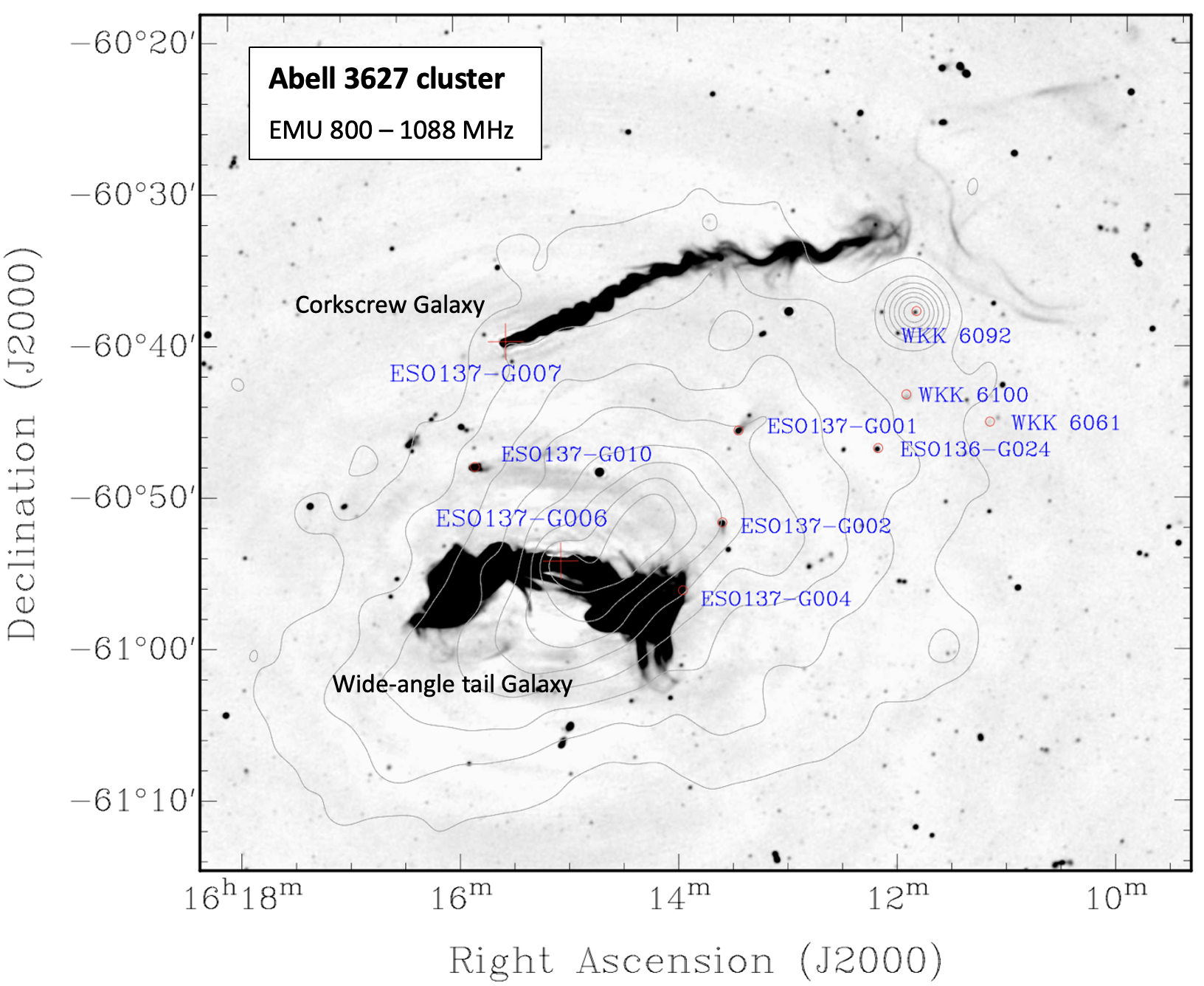}
\caption{ASKAP 944~MHz radio continuum image of galaxy cluster Abell~3627 (made using robust $r = 0.25$ weighting) overlaid with ROSAT PSPC (0.5--2~keV) X-ray contours (at $\sim$20, 30, ... 90\% of the peak flux). The two very bright and extended radio sources are the head-tail galaxy 1610--60.5, known as the Corkscrew Galaxy, and the wide-angle tail galaxy 1610--60.8. All detected cluster galaxies are labelled. The ASKAP  synthesized beam is $14\farcs8 \times 13\farcs0$.} 
\label{fig:a3627-field-labels}
\end{figure*}

Abell~3627 (A3627) -- also known as the Norma Cluster -- is a rich, nearby galaxy cluster \citep{Abell1989} at the core of the Great Attractor \citep{Dressler1987,Lynden-Bell1988, Kraan1996}. Its location near the Galactic bulge at low Galactic latitude ($l,b$ = 325$^{\circ}$, $-7^{\circ}$) causes A3627 to be partially hidden from view by foreground dust and stars. Kinematic analysis by \citet{Woudt2008}, based on $\sim$300 cataloged cluster galaxies within the 2~Mpc Abell radius, suggests a mean velocity of $4871 \pm 54$\kms\ for A3627 and a velocity dispersion of 925\kms\ \citep[see also][]{Kraan1996}. We adopt a cluster distance of 70~Mpc. The cluster centre, as determined from the peak of the X-ray distribution \citep{Boehringer1996,Tamura1998}, approximately coincides with the wide-angle tail (WAT) radio galaxy 1610--60.8 \citep{Ekers1970}, with host elliptical galaxy ESO\,137-G006 (\vhel\ = 5448\kms). The X-ray derived gas temperature of $6.7 \pm 0.3$~keV for the central region of A3627 \citep{Tamura1998} is consistent with its high velocity dispersion and among the highest known \citep[e.g., similar to A2256,][]{Rottgering1994}. Interferometric \HI\ studies in part of A3627 \citep{Vollmer2001} detect only two galaxies (WKK~6801 and WKK~6489) in the cluster outskirts, confirming the expected \HI-deficiency in such a bright X-ray cluster. 

\begin{figure*} % Figure 2
\centering
  \includegraphics[width=16cm]{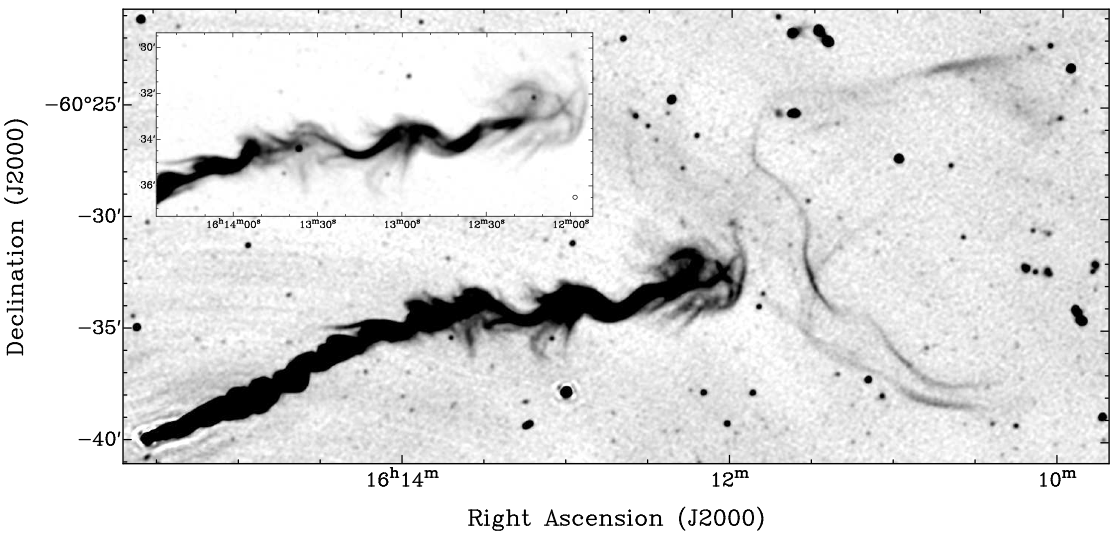}
\caption{ASKAP 944~MHz radio continuum image of the Corkscrew Galaxy (1610--60.5), made using robust $r = 0.2$ weighting. The intricate filamentary structure of the radio tail, which is shaped like a giant corkscrew on the sky, is seen in unprecedented detail. The inset shows part of the Corkscrew galaxy tail at a different greyscale contrast to highlight the wispy threads and filaments along its older part. Here the ASKAP synthesized beam is $11\farcs6 \times 10\farcs9$. --- Note that minor artefacts remain near the Corkscrew Galaxy head in the south-east corner as well as across the image. }
\label{fig:Corkscrew-grey}
\end{figure*}

Only $\sim$15\arcmin\ north of the WAT galaxy (i.e., 300~kpc projected distance), near the cluster periphery, lies the stunning head-tail (HT) "Corkscrew Galaxy" 1610--60.5, originally discovered by \citet{Ekers1969}, with the name referring to its long, highly collimated and twisting (helical) radio tail (see Figs.~\ref{fig:a3627-field-labels} \& \ref{fig:Corkscrew-grey}). Its host galaxy, ESO\,137-G007 (\vhel\ = 4945\kms), is likely moving at high speed (from west to east) through the intracluster medium (ICM) perpendicular to our line-of-sight. For a summary of the WAT and HT galaxy properties and references see Table~\ref{tab:properties}.

Both cluster radio galaxies, 1610--60.8 and 1610--60.5, were studied in detail using the Fleurs Radio Telescope at 1415~MHz \citep{Christiansen1977}, the Molonglo Observatory Synthesis Telescope (MOST) at 843~MHz \citep{JonesMcAdam1992, JonesMcAdam1994} and the Australia Telescope Compact Array (ATCA) at 1360 and 2370~MHz \citep{JonesMcAdam1996}. Recent MeerKAT observations, focused only on the WAT radio galaxy 1610--60.8, delivered stunning new 1.0~GHz and 1.4~GHz radio continuum and spectral index images with an rms of $\sim$20~$\mu$Jy\,beam$^{-1}$ at $\sim$10\arcsec\ resolution \citep{Ramatsoku2020}. Narrow synchrotron threads of unknown origin are found stretching between the two bent radio lobes, also visible in Fig.~\ref{fig:a3627-field-labels}. See \citet{Condon2021} for a study of similar threads associated with the jets of the FR\,I radio galaxy IC~4296.

The neighboring "Corkscrew Galaxy" (1610--60.5) -- the main target of this ASKAP paper -- is less powerful but much larger than the WAT radio galaxy (see Table~\ref{tab:properties}). Its long radio tail is, as the name suggests, highly collimated and twisted, consisting of a well-studied $\sim$6\arcmin-long bright inner part, gradual bending at $\sim$10\arcmin, followed by a fainter outer part with increasing amplitude wiggles. Using ATCA and MOST data, \citet{JonesMcAdam1994} examine the brightness, width and alignment of the Corkscrew tail along its full length, while \citet{JonesMcAdam1996} analyse the polarisation and spectral index along its bright inner part. 

Similar bent tail radio galaxies are found in other clusters such as Perseus \citep{Gendron-Marsolais2020,Roberts2022}, Coma \citep{Lal2022}, IIZW108 \citep{Mueller2021}, Abell~3266 \citep{Riseley2022}, Shapley \citep{Venturi2022}, Abell~2255 \citep{Botteon2020,Ignesti2023} and Abell~2256 \citep{Rottgering1994,Owen2014}. High-resolution, multi-frequency radio continuum studies of narrow angle tail (NAT) and head-tail (HT) radio galaxies, their morphologies, orientations and spectral indices, are increasing in the literature \citep[see, e.g.,][]{ODea1987,deVos2021,Morris2022}, and show, for example, the spectral index to significantly steepen from the galaxy head to its tail end due to the progressive electron ageing along the jet \citep[e.g.,][]{Mueller2021,Cuciti2018}. The radio tails are shaped by their motions through and interactions with a dynamically active ICM (e.g.,  turbulence, sloshing, winds and shocks). See \citet{Nolting2019, Nolting-II2019} for 3D magneto-hydrodynamical (MHD) simulations of interactions between radio galaxies and the ICM, in particular the shock impacts on active radio jets. \\

\begin{table*}
\centering
\begin{tabular}{cccc}
\hline
Properties & wide-angle tail (WAT) & head-tail (HT) & References \\
 & radio galaxy & radio galaxy \\
\hline
\hline
names & 1610--60.8 & 1610--60.5 & \citet{Christiansen1977} \\
 & B1610--608 & B1610--605 & \citet{JonesMcAdam1996} \\
 & & "Corkscrew Galaxy" \\
host galaxy & ESO\,137-G006 & ESO\,137-G007 & \citet{ESO-LV1989} \\
 & WKK~6269 & WKK~6305 & \citet{Woudt2001} \\ 
 & 2MASS J16150381--6054258 & 2MASS J16153291--6039552 & \citet{Skrutskie2006} \\
 & 2XMM J161503.8--605425 & 2XMM J161533.1--603953 & \citet{Watson2009} \\
redshift & 0.0176 & 0.0162 & \citet{Whiteoak1972,Christiansen1977} \\
% redshifts: I.J. Danziger, priv comm.
 & 0.0182 & 0.0165 & \citet{Woudt2008} \\
velocity & $5448 \pm 35$\kms & $4945 \pm 35$\kms & \citet{Woudt2008} \\ 
% heliocentric
peculiar velocity & $\sim$577\kms & $\sim$74\kms & \citet{Woudt2008} \\
type & E1 & S0 & \citet{RC3} \\
1410~MHz flux & 45.4 Jy & 5 Jy & \citet{Ekers1970} \\
843~MHz flux  & 90.0 Jy & 6 Jy & \citet{JonesMcAdam1992} \\
843~MHz size, $PA$  & 25$'$, 112\degr & 16$'$, 88\degr & \citet{JonesMcAdam1992} \\
% 408~MHz flux  &   & 6.1 Jy & Large et al. 1981 \\
% 4850 MHz flux &   & 1.35 Jy & PMN \\
\hline
\end{tabular}
\caption{Properties of the two large radio galaxies in the Abell~3627 cluster and their optical host galaxies from the literature. The quoted peculiar velocity is with respect to the mean velocity of the cluster galaxies \citep[$4871 \pm 54$\kms,][]{Woudt2008}.}
\label{tab:properties}
\end{table*}

In this paper we focus on the Corkscrew Galaxy and its giant radio tail. The re-processing of data from ASKAP survey observations at 944~MHz and 1.4~GHz are described in Section~2. Our results are presented in Section~3, followed by the discussion in Section~4, where we consider possible formation mechanisms for the intricate filamenatry structure along and beyond the Corkscrew tail. Our conclusions are given in Section~5.

\section{ASKAP Observations and Data Processing}
\label{sec:obs}

The Australian Square Kilometer Array Pathfinder \citep[ASKAP,][]{Johnston2008} is a radio interferometer located in the Murchison Radio Observatory (MRO), consisting of $36 \times 12$-m antennas with baselines up to 6.4~km in extent. The resulting high angular resolution is complemented by good surface-brightness sensitivity thanks to the dense core of 30 antennas within a 2-km diameter area. Each antenna is equipped with a wide-field Phased Array Feed (PAF), operating at frequencies from 700~MHz to 1.8~GHz \citep{Chippendale2015}. For a comprehensive ASKAP overview see \citet{Hotan2021}. A selection of ASKAP science highlights is presented in \citet{Koribalski2022}. 

For this project we use ASKAP data centred at 944~MHz and 1.4~GHz from the EMU \citep{EMU,EMU-PS} and WALLABY \citep{Koribalski2012,Koribalski2020} survey projects, respectively. The 2020 WALLABY pilot field targeting the Norma cluster was centred on $\alpha,\delta$(J2000) = $16^{\rm h}\,16^{\rm m}\,35.8^{\rm s}$, $-59\degr\,29\arcmin\,15\arcsec$ and a bandwidth of 144~MHz was used to avoid known radio interference. The 2023 EMU main survey field was centred on $\alpha,\delta$(J2000) = $16^{\rm h}\,27^{\rm m}\,25.7^{\rm s}$, $-60\degr\,19\arcmin\,18\arcsec$, using the full 288~MHz bandwidth. The pipeline-calibrated multi-channel continuum visibilities of both fields were re-processed (see below) to improve the image quality and dynamic range around the two very bright and extended radio sources in A3627. See Table~\ref{tab:obs} for a summary of the observations and image properties.

\begin{table}
\centering
\begin{tabular}{ccc}
\hline
ASKAP  & SB~53218 & SB~11816 \\
\hline
\hline
date   & 18 Sep 2023 & 14 Feb 2020 \\
integration time & 10~h & 8~h \\
centre frequency & 943.5~MHz & 1367.5~MHz \\
bandwidth        & 288~MHz & 144~MHz \\
field of view    & $\sim$30 deg$^2$ 
                 & $\sim$30 deg$^2$ \\
beam footprint   & closepack & square$\_$6x6 \\
angular resolution & \\
~~ uniform & $6\farcs4 \times 6\farcs1$ & $5\farcs2 \times 4\farcs6$ \\
~~ uniform+taper & $15\farcs1 \times 15\farcs0$ & $15\farcs0 \times 15\farcs0$\\
~~ robust $r = 0$ & $11\farcs6 \times 10\farcs9$ & $8\farcs8 \times 8\farcs2$ \\
~~ robust $r = 0.25$ & $14\farcs8 \times 13\farcs0$ & $10\farcs0 \times 9\farcs6$ \\
mean rms noise & \\
~~ uniform & 104~$\mu$Jy\,beam$^{-1}$ & 53~$\mu$Jy\,beam$^{-1}$ \\
~~ uniform+taper & 58~$\mu$Jy\,beam$^{-1}$ & 50~$\mu$Jy\,beam$^{-1}$ \\
~~ robust $r = 0$ & 40~$\mu$Jy\,beam$^{-1}$ 
    & 30~$\mu$Jy\,beam$^{-1}$ \\ 
~~ robust $r = 0.25$ & 51~$\mu$Jy\,beam$^{-1}$ 
    & 30~$\mu$Jy\,beam$^{-1}$ \\ 
\hline
\end{tabular}
\caption{ASKAP wide-field observations and image properties.}
\label{tab:obs}
\end{table}

\subsection{ASKAP SB~53218}

The 944~MHz ASKAP images available on CASDA for SB~53218 suffer (a) residual $w$-term artefacts and (b) some RFI towards the end of the observation. To improve their quality, we re-image a subset of the 36 PAF beams (1, 6, 7, 12, 13, 14, 18, 19, and 25) using the pipeline described by \citet{Duchesne2024}. This process includes flagging baselines $>$200~m for the last 3.5~hours of the observation. This had a negligible effect on the overall sensitivity, but significantly reduces artefacts around the bright core of the Corkscrew Galaxy. After flagging, we extract the region around Abell~3627 by subtracting a model of the visibilities away from the cluster and performing additional phase and amplitude self-calibrating on the extracted dataset in the direction of the cluster. We use \emph{wsclean} \citep{Offringa2014,Offringa2017} and \emph{casa} \citep{CASA2022} for imaging and calibration tasks, then create three final sets of images for each beam: with uniform weighting, with `Briggs' \citep{Briggs95} robust $r = 0.0$ and $r = +0.25$ weighting. 

Other ASKAP surveys have highlighted small variations in the relative astrometry of sources between PAF beams \citep{McConnell2020,Duchesne2023}. We attempt to correct for this to ensure relative astrometry between ASKAP images is consistent for the uniformly-weighted (and highest-resolution) images. We use large template images of the full PAF beam to generate the RA and declination offsets. We use the \emph{aegean} \citep{aegean1,aegean2} source-finder to generate per-beam source lists, then match these individual source lists to an equivalent source list of full field image available through CASDA. We reject sources without a match within $\approx $10\arcsec, with ratios of integrated ($S_\text{int}$) to peak ($S_\text{peak}$) flux density of $S_\text{int}/S_\text{peak}>1.2$, and any sources with neighbours within 36\arcsec\ in either catalogue. A median offset in both RA and DEC is then calculated for each PAF beam after rejecting outliers and applied to the FITS image headers. The median offsets per beam range from --0\farcs33 to +0\farcs60 in RA and --0\farcs01 to +0\farcs40 in DEC, with standard deviations of $\sim$0\farcs8. The astrometry corrections are only applied to the uniformly-weighted wide-band image.

The individual beam images are then combined as a linear mosaic, weighted by the primary beam response. The primary beam model is the same as used in the original pipeline processing, derived from holography observations. We also make a separate set of uniform images with a minimum $(u,v)$ cut of 110$\lambda$, corresponding to the minimum $(u,v)$ value of the 1.4~GHz data (at the top of the band). This additional uniform image is also imaged with a 15\arcsec\ Gaussian taper to the visibilities. We measure rms noise values of 104, 58, 40 and 51~$\mu$Jy\,beam$^{-1}$, for the uniform, uniform+taper, and $r = 0.0$, and $r = +0.25$ linear mosaic images, respectively, near the Corkscrew Galaxy. The following resolutions are achieved for the wide-band images for uniform weighting ($6\farcs4 \times 6\farcs1$, $PA = 27\degr$), {uniform + 15\arcsec\ taper }($15\farcs1 \times 15\farcs0$, $PA = -37\degr$), robust $r = 0$ ($11\farcs6 \times 10\farcs9$, $PA = -124\degr$), and robust $r = +0.25$ ($14\farcs8 \times 13\farcs0$, $PA = +116\degr$). 

In addition to the ASKAP 944~MHz wide-band image of A3627, we generate 16 sub-band images of 18~MHz width. Spectral index images are created after all sub-band images are convolved to a common resolution at the respective weighting. Furthermore, we create stacks of 4 sub-band images, each with width of 72~MHz. Note that the sub-band images have slightly lower angular resolution than the respective wide-band images.

\subsection{ASKAP SB~11816}

We perform similar re-imaging of the 1.4~GHz ASKAP datasets as the CASDA images for SB~11816 suffer many artefacts. ASKAP data in this band typically have the lower half of the band flagged due to persistent RFI, and in this case the total bandwidth is 144~MHz. Due to a different pointing direction and primary beam size, we image a slightly different subset of PAF beams (2, 3, 11, 12, 13, 14). For imaging and self-calibration, we image the full PAF beam as opposed to just a small region around Abell~3627. As the archival images are of fairly poor quality for the 1.4~GHz data, self-calibration is performed with more loops and a significantly more gradual lowering of the CLEAN threshold for each loop to more carefully avoid artefacts in the self-calibration model. Finally, as SB~11816 does not have a holography-derived primary beam model associated with it, we opt to assume a circular Gaussian model which generally performs well in the centre of the ASKAP field and gets worse for edge/corner PAF beams. Linear mosaicking of individual beams is the same as for the 944~MHz data, though because of the smaller fractional bandwidth ($\approx$10\%) we do not use the sub-band images created during the usual imaging process.

We repeat the astrometric correction procedure that was performed for the 944~MHz dataset, finding a wider range of offsets: between --0.05\arcsec\ to +2.75\arcsec\ in RA and +0.15\arcsec\ to +1.68\arcsec\ in DEC. The pixel size for the tapered images is 2\farcs5. Linear mosaics of all beams are formed similarly to the 944-MHz data. The mosaic images have angular resolutions of, $5\farcs2 \times 4\farcs6, PA = 58.1\degr$ (uniform) 15\arcsec\ (uniform + taper), $8\farcs8 \times 8\farcs2, PA = 60.7\degr$ (robust $r = 0.0$), and $10\farcs0 \times 9\farcs6, PA = 52.8\degr$ (robust $r = +0.25$) images, and the mean rms noise measured near the Corkscrew Galaxy is 53, 50, 30, and 30~$\mu$Jy\,beam$^{-1}$, respectively.

\begin{figure} % Figure 3
\centering
   \includegraphics[width=8.5cm]{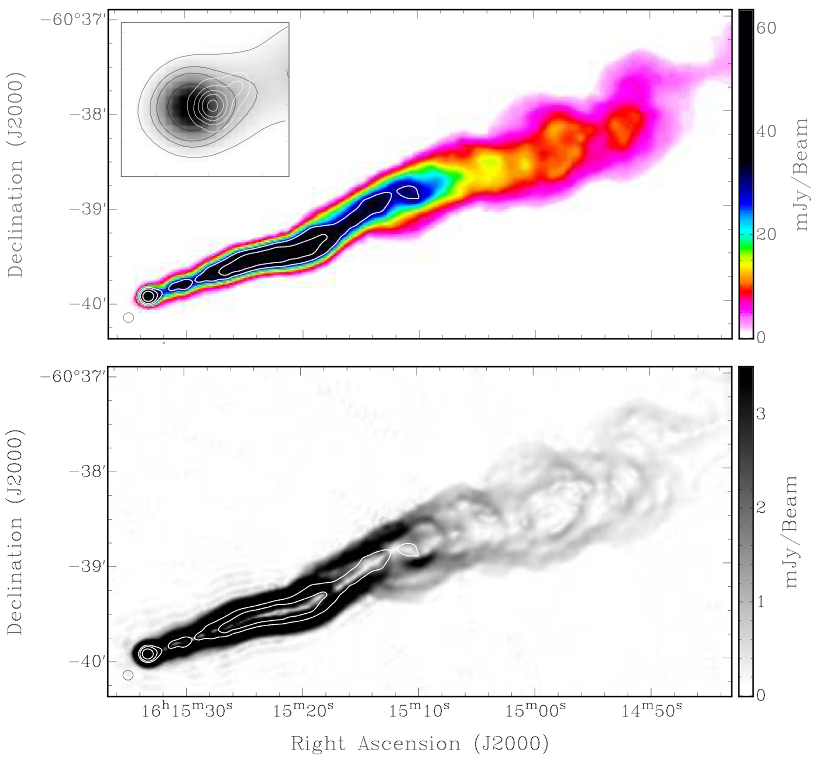}
\caption{High-resolution ASKAP 944~MHz images of the Corkscrew Galaxy's bright inner region. {\bf -- Top:} Radio continuum emission (white contours at 30, 50, and 100~mJy\,beam$^{-1}$). The {\bf inset} (size = 20\arcsec) shows the eastern-most part of the radio emission (black contours at 10, 30, 50, 100 and 150~mJy\,beam$^{-1}$) near the host galaxy ESO\,137-G007 (white contours from 2MASS $K$-band). The radio peak is offset to the east by $3\arcsec \pm 1\arcsec$, possibly hinting at the emerging counter-jet. {\bf -- Bottom:} As above after applying the miriad task {\em imsharp} to highlight fine-scale structure in the radio tail. The ASKAP synthesized beam ($6\farcs4 \times 6\farcs1$) is shown in the bottom left corner of each panel.} 
\label{fig:corkscrew-bright}
\end{figure}

\subsection{ASKAP spectral index maps}

\begin{figure*} % Figure 4
\centering
    \includegraphics[width=0.8\linewidth]{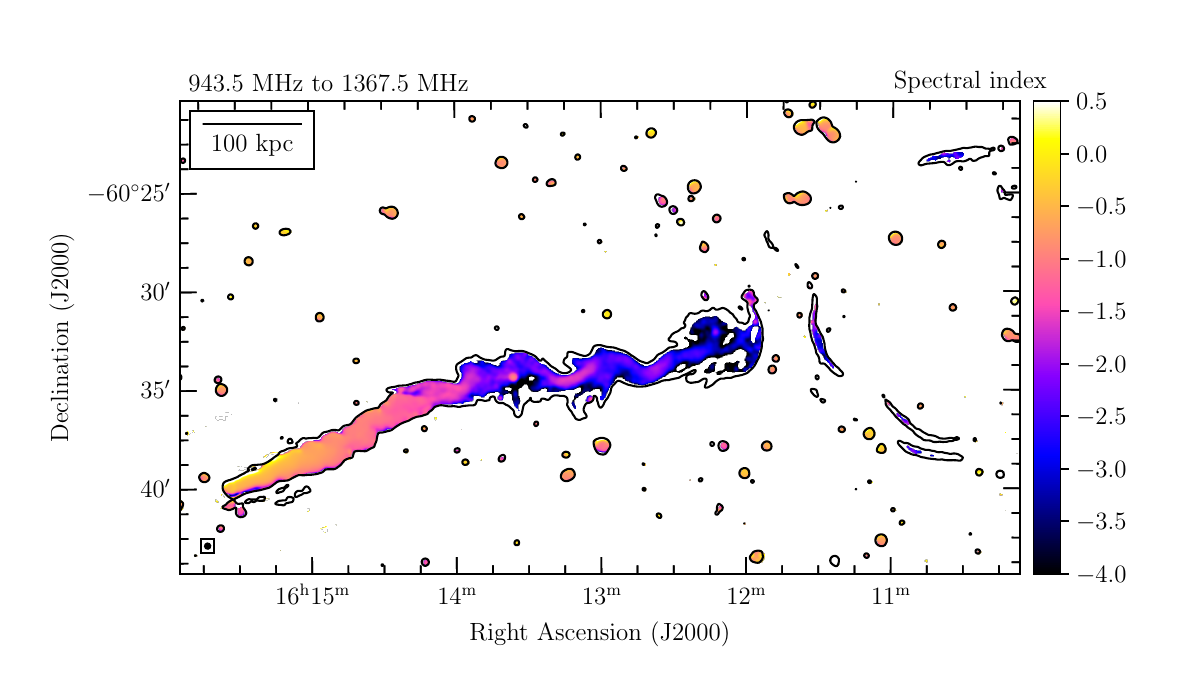}\\
    \includegraphics[width=0.8\linewidth]{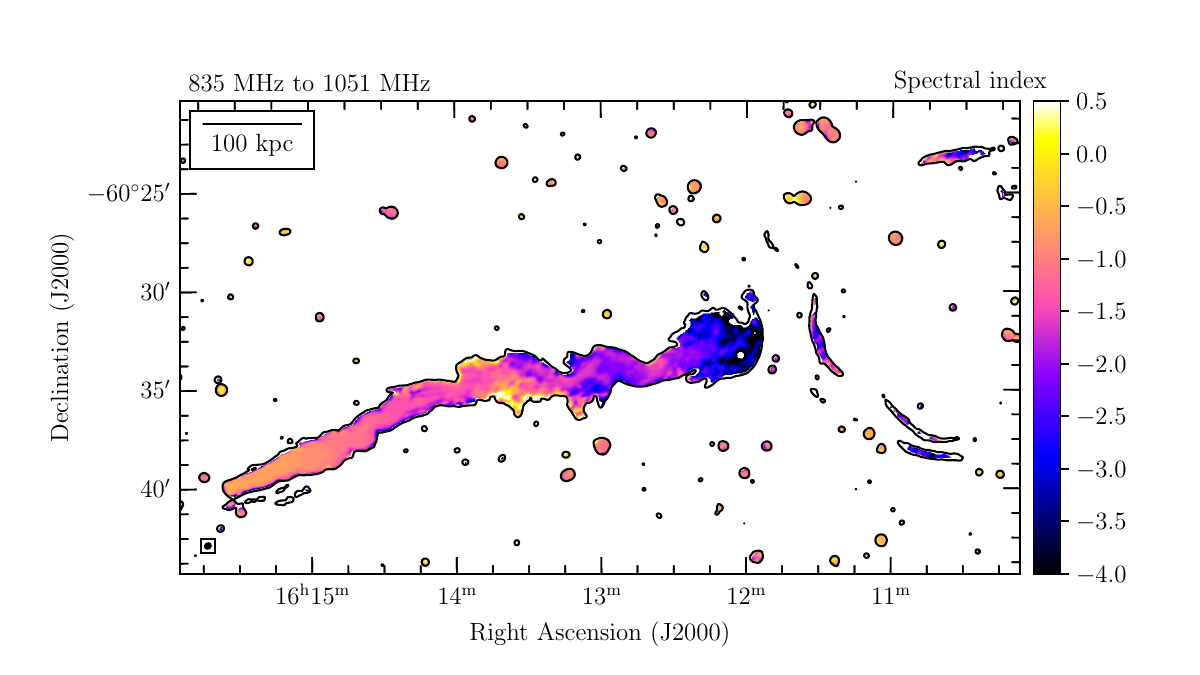}
\caption{\label{fig:spindex:cork} Spectral index maps of the Corkscrew Galaxy between 944 -- 1368~MHz (top), as derived by using the ASKAP images from both the EMU and WALLABY projects, and between 835 -- 1051~MHz (bottom), derived using only the EMU data. The black contour is drawn at $5 \sigma_\text{rms}$ {($\sigma_\text{rms} = 64$\,$\mu$Jy\,beam$^{-1}$)} of the 944~MHz image used for the spectral index calculation. The beams (top: $16\farcs0 \times 16\farcs0$; bottom: $18\farcs4 \times 15\farcs9, PA = 124.7\degr$) are shown in the bottom left corner of each panel.}
\end{figure*}

\begin{figure} % Figure 5
\centering
    \includegraphics[width=1\linewidth]{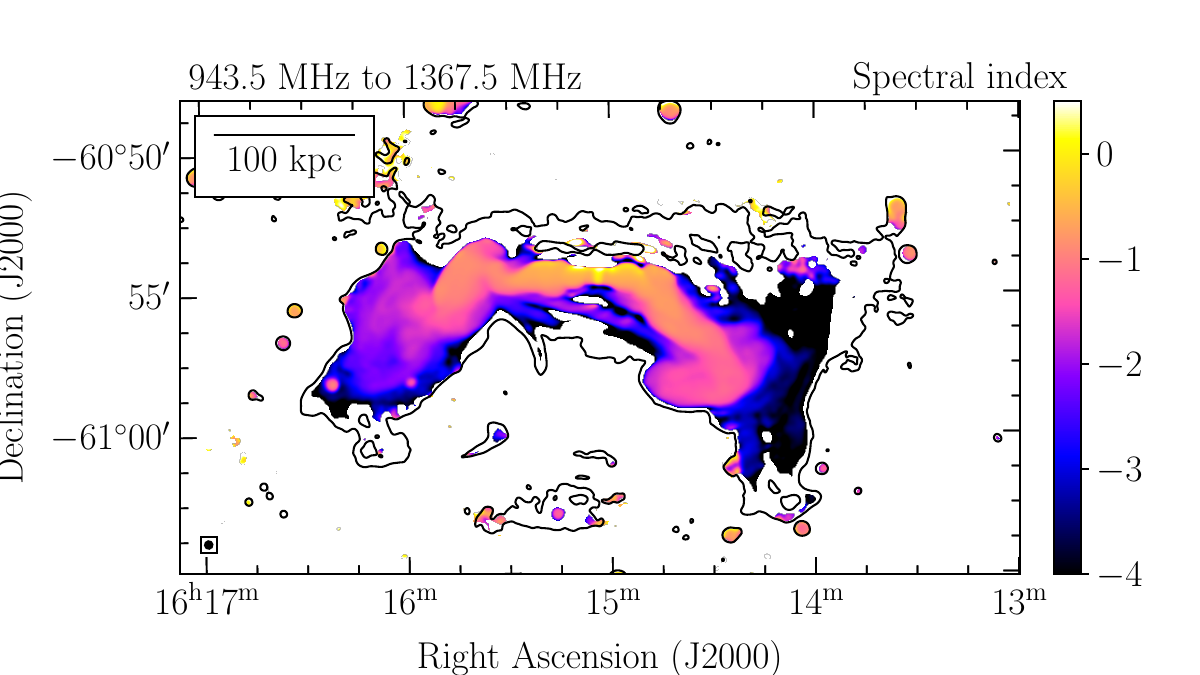}\\
    \includegraphics[width=1\linewidth]{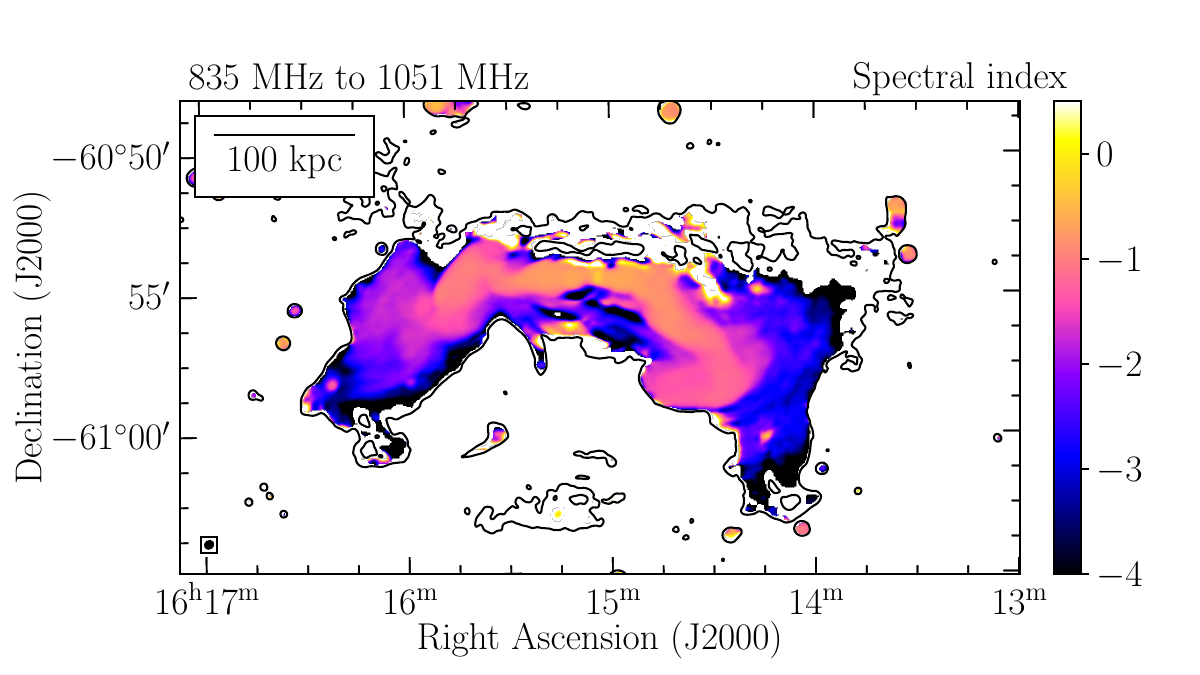}
\caption{\label{fig:spindex:wat} Spectral index map of the WAT between 944 -- 1368~MHz (top) and 835 -- 1051~MHz (bottom). The black contour is drawn at $5 \sigma_\text{rms}${($\sigma_\text{rms} = 64$\,$\mu$Jy\,beam$^{-1}$)} of the 944~MHz image used for the spectral index calculation. The beams (top: $16\farcs0 \times 16\farcs0$; bottom: $18\farcs4 \times 15\farcs9, PA = 124.7\degr$) are shown in the bottom left corner of each panel.}
\end{figure}

We make two spectral index ($\alpha$) maps for each of the Corkscrew and WAT radio galaxies, shown in Figs.~\ref{fig:spindex:cork} \& \ref{fig:spindex:wat}, respectively. 
\begin{itemize}
    \item An EMU in-band $\alpha$ map (835 -- 1051~MHz), created using the uniformly weighted, common-$\lambda$ 72~MHz sub-band images at $18\farcs4 \times 15\farcs9$ resolution. For all pixels above a position-dependent $5\sigma_\text{rms}$, we fit a power law to obtain the spectral index.
    \item A two-point $\alpha$ map (944 -- 1368~MHz) using both the EMU and WALLABY datasets at $16\arcsec \times 16\arcsec$ resolution, created as follows,
    \begin{equation}
    ~~~~~~~~~~~~~~~~~~~\alpha = \frac{\log_\text{10}\left(S_\text{943.5} / S_\text{1367.5}\right)}{\log_\text{10}\left(943.5 / 1367.5\right)} \, , 
    \end{equation}  
    using only pixels above a position-dependent $5\sigma_\text{rms}$.
\end{itemize}

\section{Results} % Section 3

Figure~\ref{fig:a3627-field-labels} shows the re-processed ASKAP 944~MHz radio continuum image of the A3627 cluster. The northern part of the image reveals the very long, slightly curved, filamentary tail of the Corkscrew Galaxy (1610--60.5) in unprecedented detail. The Corkscrew tail is surprisingly straight and collimated before slightly changing direction. Figure~\ref{fig:Corkscrew-grey} highlights the intricate filamentary structures within the tail which increase in complexity and width towards its western X-shaped end. Further west and disconnected from the Corkscrew tail (but most likely part of it), we find a set of very narrow, arc-shaped filaments (threads), most of which are oriented roughly perpendicular to the tail direction. The southern part of Fig.~\ref{fig:a3627-field-labels} is dominated by the WAT radio galaxy (1610--60.8) whose bright lobes show numerous synchrotron threads including several connecting the bent inner lobes, as already shown by \citet{Ramatsoku2020} using MeerKAT data. In between the two bright radio galaxies we detect numerous, much smaller spiral galaxies, some with short radio tails. \\

Overlaid onto the ASKAP radio continuum image of Abell~3627 in Fig.~\ref{fig:a3627-field-labels} are X-ray contours from a ROSAT PSPC (0.5--2~keV) image \citep[see][]{Boehringer1996} smoothed to a resolution of 150\arcsec. The observed X-ray substructure, as discussed by \citet{Boehringer1996}, \citet{Tamura1998} and \citet{Nishino2012}, is indicative of a cluster merger. While the Corkscrew Galaxy is located near the northern cluster periphery, the WAT galaxy appears to be located near or approaching the cluster centre. The orbital motion of the Corkscrew Galaxy proceeds from west to east, approximately perpendicular to the line-of-sight, most likely heading around the cluster potential \citep[$M_{\rm 500} = 2.4 \times 10^{14}$\Msun,][]{Planck2016} which is traced by the bright and extended X-ray emission. The velocity of the Corkscrew Galaxy is likely similar to that of the WAT radio galaxy, which has a line-of-sight peculiar velocity of $\sim$600\kms\ (see Table~\ref{tab:properties}). Interestingly, the set of thin, arc-shaped synchrotron filaments in the western periphery of the A3627 cluster is found where the X-ray emission has decreased to less than 20\% peak brightness. The bright X-ray point source near the tail end likely originates from the Seyfert\,1 galaxy WKK~6092. \\

In the following we examine the morphology and spectral index of the Corkscrew Galaxy tail (Section~3.1), its intricate filamentary structure (Section~3.2), the set of arc-shaped radio filaments west of the tail (Section~3.3), and the small jellyfish galaxies detected in the cluster (Section~3.4).

\subsection{Corkscrew Tail Morphology and Spectral Index} % Section 3.1

\begin{figure*} % Figure 6
\centering
  \includegraphics[width=16cm]{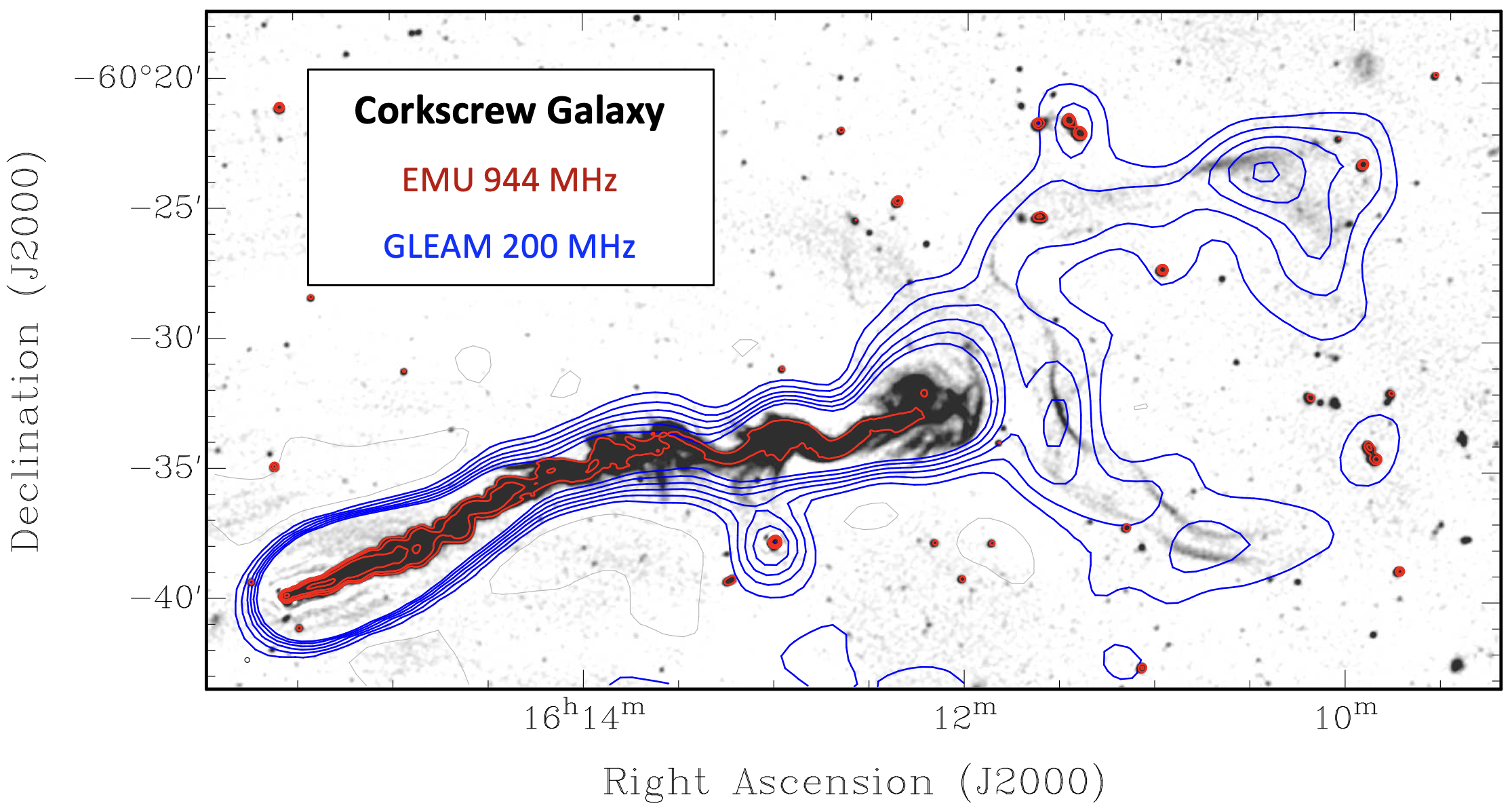}
\caption{ASKAP 944~MHz wide-band image of the Corkscrew Galaxy overlaid with GLEAM 170 -- 231~MHz contours. The ASKAP contours are in red (1.5, 6, 30, and 150~mJy\,beam$^{-1}$); the GLEAM contours are in blue (0.1, 0.2, 0.3, 0.4, 0.5 and 0.6 mJy\,beam$^{-1}$) and grey (--0.1 mJy\,beam$^{-1}$). The ASKAP synthesized beam ($11\farcs6 \times 10\farcs9$ for $r = 0$) is shown in the bottom left corner. The set of western arc-shaped filaments form a bubble-like extension of the collimated Corkscrew tail with steep spectral index (see Section 3.4).}
\label{fig:corkscrew-emu+gleam}
\end{figure*}

We measure a Corkscrew tail length of $\sim$28\arcmin, from the radio peak in the east to the faint X-shaped feature in the west (see Fig.~\ref{fig:Corkscrew-grey}). For the adopted cluster distance of 70~Mpc this length corresponds to $\sim$570~kpc. If we include the western set of filaments beyond the X-shaped tail end, the total length becomes $\sim$45\arcmin\ or 920~kpc. The radio tail is also detected in the low-resolution GLEAM images \citep{Wayth2015,Hurley-Walker2017}, while only the bright, inner part of the tail (see Fig.~\ref{fig:corkscrew-bright}) is detected in the old ATCA 2.3~GHz image \citep[][project C180]{JonesMcAdam1996}. Spectral index maps of the Corkscrew and the WAT radio galaxies are presented in Figs.~\ref{fig:spindex:cork} and \ref{fig:spindex:wat}, respectively. Furthermore, in Figs.~\ref{fig:corkscrew-emu+gleam} and \ref{fig:corkscrew-rgb} we show the EMU 944~MHz wide-band image overlaid with GLEAM 170--231~MHz contours to highlight the western set of arc-shaped filaments and their location with respect to the X-ray emission; the GLEAM resolution at that frequency is $\sim$2\farcm5. For a discussion of the likely formation mechanisms see Section~4. \\

Fig.~\ref{fig:corkscrew-bright} shows the well-known inner, bright part of the Corkscrew Galaxy tail which consists of a highly collimated jet emerging from near the central black hole in the lenticular galaxy ESO\,137-G007. The narrow radio jet, which is unresolved at the base, initially broadens to a width of $\sim$60\arcsec\ (20~kpc) at 6\arcmin\ length before re-collimating. Its fine-scale structure is emphasized in the sharpened image. Radio emission slightly offset to the east from the optical galaxy, could possibly be an emerging counter-jet that is bending backwards, merging with the western tail \citep[see also][]{JonesMcAdam1996}. No sign of two jets merging to form the western tail is detected at the current resolution. High-frequency radio images at sub-arcsec resolution may be needed to detect such a twin jet \citep[see, e.g.,][]{deGregory2017}. \\

Mild oscillations of the Corkscrew tail are already noticeable in the inner tail, starting at $\sim$1\arcmin\ distance from the core (see Fig.~\ref{fig:corkscrew-bright}) and growing substantially towards the X-shaped tail end. There is delicate balance between the jet pressure which decreases away from the AGN and the surrounding ICM pressure, causing the re-collimation with a characteristic scale length \citep[see][and references therein]{Velovic2022}. These semi-regular oscillations, which were already noted and characterised by 
\citet{JonesMcAdam1994,JonesMcAdam1996}, get larger towards the tail end. We measure a maximum oscillation amplitude of $\sim$2\arcmin\ or 40~kpc in the old part of the tail.  \\

About half way through its full length (at $\sim$14\arcmin\ or 285~kpc distance from the core) the highly collimated Corkscrew tail appears disrupted and gradually changes direction by $\sim$20\degr. The zoomed-in ASKAP 944~MHz image in Fig.~\ref{fig:Corkscrew-grey} highlights the intricate filamentary structure in this tail section (see also Section~3.2). Generally, from midway on, the tail -- as it ages -- becomes more filamentary and decreases in brightness, while the amplitude and period of the oscillations grow. The change occurs where (a) the jet seems to decollimate (synchrotron emission almost disappears), and (b) there is a ridge of brighter X-ray emission heading almost radially outwards in the cluster (see Fig.~\ref{fig:a3627-field-labels}). Deeper X-ray images of that area are needed to confirm this. \\

The X-shaped feature and surrounding filaments at the end of the connected Corkscrew tail are clearly seen in Fig.~\ref{fig:Corkscrew-grey}. The filaments, which vary in length from $\sim$1\arcmin\ -- 4\arcmin\ ($\sim$20 -- 80~kpc), are somewhat similar to the set of arc-shaped filaments west of the collimated tail end (see Section~3.4). Fig.~\ref{fig:spindex:cork} highlights their comparable spectral indices. Deeper, high-resolution radio continuum images are likely to reveal further filamentary structure around the X-shaped feature at the tail end, only hinted at in our images.  \\

Fig.~\ref{fig:spindex:cork} shows the gradual steepening of the spectral index along the Corkscrew Galaxy tail. Using the EMU in-band spectral index map, we measure $\alpha = -0.56$ near the radio core to $\alpha = -4$ at the X-shaped tail end. The spectral index gradient along the Corkscrew's bright inner tail was previously noted by \citet{JonesMcAdam1996}. The steepening of the spectrum with distance from the host galaxy is due to synchrotron aging of the radio-emitting plasma. Similar gradients are found along the radio lobes of the WAT radio galaxy (see Fig.~\ref{fig:spindex:wat}), studies in detail by \citet{Ramatsoku2020}.

\subsection{Corkscrew Tail Filamentary Structure} % Section 3.2

The Corkscrew tail starts showing complex filamentary structure from $\sim$10\arcmin\ length. The extent of these filaments, which consist of spurs, arcs and wisps, grows in the fainter part of the oscillating tail (see Fig.~\ref{fig:Corkscrew-grey}). And some resemble those in the neighbouring WAT 1610--60.8 (see Fig.~\ref{fig:a3627-field-labels} and discussion). Narrow filaments (length $\sim$ 1\arcmin\ -- 4\arcmin\ or $\sim$20 -- 80~kpc) are seen at various angles, including many that are forward facing, i.e. in the direction of the galaxy motion. They appear to grow in length and complexity towards the tail end, likely shaped by the dynamics of the surrounding ICM.

\subsection{The western set of arc-shaped radio filaments} % Section 3.3

Beyond the Corkscrew's X-shaped tail end in the western periphery of the A3627 cluster, a disconnected set of very thin, arc-shaped synchrotron threads or filaments are detected, forming a partial bubble of $\sim$15\arcmin\ diameter (see Fig.~\ref{fig:Corkscrew-grey}). While the western filaments are similar to those at the X-shaped end of Corkscrew tail, which suggests both are related, they are typically longer ($\sim$5\arcmin\ or 100~kpc) and thinner than the filaments at the collimated tail end. Most of these filaments are aligned roughly perpendicular to the tail, forming a double-wave structure. This group of western filaments look like the rim of an expanding bubble. Interestingly, \citet{Nolting2019} see such structure in their simulations which we discuss in Section~4. Fig.~\ref{fig:a3627-field-labels} shows this bubble to lie just outside the lowest X-ray contour. We have spectral index measurements ($\alpha \approx -2.5 \pm 0.5$) for four of the western radio filaments seen in Fig.~\ref{fig:spindex:cork}), i.e. similar to the values at the tail end. Collectively these filaments are detected in the low-frequency GLEAM maps (see Fig.~\ref{fig:corkscrew-emu+gleam}), as expected due to their steep spectral indices. We measure a GLEAM 170 -- 231~MHz flux density of $5.5 \pm 0.1$~Jy for the western extension of the tail, ie beyond the X-shaped feature at the Corkscrew tail, and an ASKAP 944~MHz flux density of $0.19 \pm 0.01$ Jy. The resulting spectral index estimate is $-2.2 \pm 0.1$.
% Spectral index estimate
% alpha = log (S1/S2) / log (nu1/nu2)
%       = log(5.50/0.19) / log (200/944)
%       = -2.2 +- 0.1

\subsection{Jellyfish Galaxies} % Section 3.4

We detect four spiral galaxies with short, radio tails ($\sim$20--30~kpc, see Table~\ref{tab:jellyfish}) -- likely caused by ram-pressure stripping and commonly referred to as jellyfish galaxies -- in Abell~3627: these are ESO\,137-G001, ESO\,137-G002, ESO\,137-G010 (see Fig.~\ref{fig:a3627-field-labels}), and WKK~6489 (outside the displayed field). Of these, only one galaxy (ESO\,137-G010) was not already known to be a jellyfish galaxy. They are very similar to jellyfish galaxies in other clusters such as IIZW108 \citep{Mueller2021}, Shapley \citep[][]{Venturi2022} and Perseus \citep{Roberts2022}, and serve as probes of shear motions in the ICM.

ESO\,137-G001, which is one of the nearest known and well-studied jellyfish galaxies, has both a $\sim$70~kpc long X-ray and a $\sim$40~kpc H$\alpha$ tail, both pointing roughly north-west, away from the cluster centre \citep{Sun2006,Sun2007,Woudt2008,Fumagalli2014}. The $\sim$30~kpc long radio tail discovered here is aligned with the bright part of the main X-ray tail. ESO\,137-G002 has a known $\sim$40~kpc long X-ray and a $\sim$20~kpc H$\alpha$ tail \citep{Sun2010,Laudari2022}, both roughly pointing south. The $\sim$20~kpc long radio tail discovered here is aligned with the above. For ESO\,137-G010 we find a $\sim$30~kpc long radio tail pointing west. The WKK~6489 radio tail has the same direction (SE) as its \HI\ tail discovered by \citet[][their Fig.~7]{Vollmer2001}.

\begin{table}
\centering
\begin{tabular}{ccccc}
\hline
 Galaxy & & systemic & \multicolumn{2}{c}{radio tail} \\
 name  & type & velocity & length & direction \\
 & & [ \kkms ] & [ $'$ / kpc ] \\
\hline
\hline
ESO\,137-G001 & SBc?  & $4461 \pm 46$ & $\sim$1.5 / 30 & NW \\ 
ESO\,137-G002 & S0?   & $5691 \pm 20$ & $\sim$1.0 / 20 & S \\
ESO\,137-G010 & SAB0  & $3378 \pm 32$ & $\sim$1.5 / 30 & W \\
WKK~6489      & Sb-Sd & $3794 \pm 58$ & $\sim$1.5 / 30 & SE \\
\hline
\end{tabular}
\caption{Properties of the radio tails of four spiral galaxies in Abell~3627. Galaxy morphologies and systemic velocities are from NED. For reference, the mean velocity of the catalogued cluster galaxies is $4871 \pm 54$\kms\ and the velocity dispersion is 925\kms\ \citep{Woudt2008}.}
\label{tab:jellyfish}
\end{table}

\section{Discussion}

The largest known HT galaxies, similar in length to the Corkscrew Galaxy, are IC\,711 in Abell~1314 \citep{Sebastian2017,Wilber2019,Srivastava2020} and 1712+638 ("the Beaver") in Abell~2255 \citep{Botteon2020,Ignesti2023}, with tail lengths of $\sim$1~Mpc. Somewhat shorter are the two HT galaxies, T1 ($\sim$550~kpc) and T2 ($\sim$400~kpc), in Abell~2142 \citep{Venturi2017,Bruno2023}. Most notable, T1 shows an arc-shaped structure beyond and perpendicular to its tail end \citep{Bruno2023}, somewhat similar to the set of western filaments found here beyond the collimated tail of the Corkscrew Galaxy. These filaments resemble those in 3D magneto-hydrodynamical simulations by \citet{Nolting2019} who explore interactions between shocks and HTs. The filamentary complex west of the X-shaped tail end may be the remnant of a vortex ring \citep[][e.g., their Fig.~6]{Nolting2019} created in a head-on collision of the Corkscrew galaxy with a shock / cold front in the cluster outskirts. The stripped plasma tails of such radio galaxies enrich the ICM  with relativistic electrons \citep[e.g.,][]{Vazza2021,Vazza2023,VazzaBotteon2024}, and may later be re-accelerated to form large-scale radio relics -- created via outwards moving merger shocks -- in the cluster outskirts \citep[e.g.,][]{vanWeeren2019}. No relics have so far been detected in A3627. \\

Studies of the X-ray emission in galaxy clusters can reveal their dynamics and evolutionary state \citep{Forman1982,Sarazin1986}. While several X-ray observations of Abell~3627 exist \citep[e.g.,][]{Boehringer1996,Sun2010,Nishino2012}, much deeper, wide-field images are needed to study the cluster outskirts. Our Figure~\ref{fig:a3627-field-labels} shows the ROSAT PSPC (0.5--2~keV) X-ray contours overlaid onto the ASKAP 944~MHz radio continuum emission of Abell~3627. Using the ROSAT data, \citet{Boehringer1996} find the X-ray emission to extend over almost one degree in radius. They find the structure to be arrow-shaped, elongated ($PA \sim 130\degr$) and asymmetric, with clear indications of an ongoing cluster merger. This is supported by two patches of excess X-ray emission, north-west and south-east of the cluster centre, when subtracting a spherically symmetric model which may indicate sub-clusters in the process of merging with the main body \citep[see also][]{Nishino2012}. \\

\begin{figure*} % Figure 7
\centering
    \includegraphics[width=16cm]{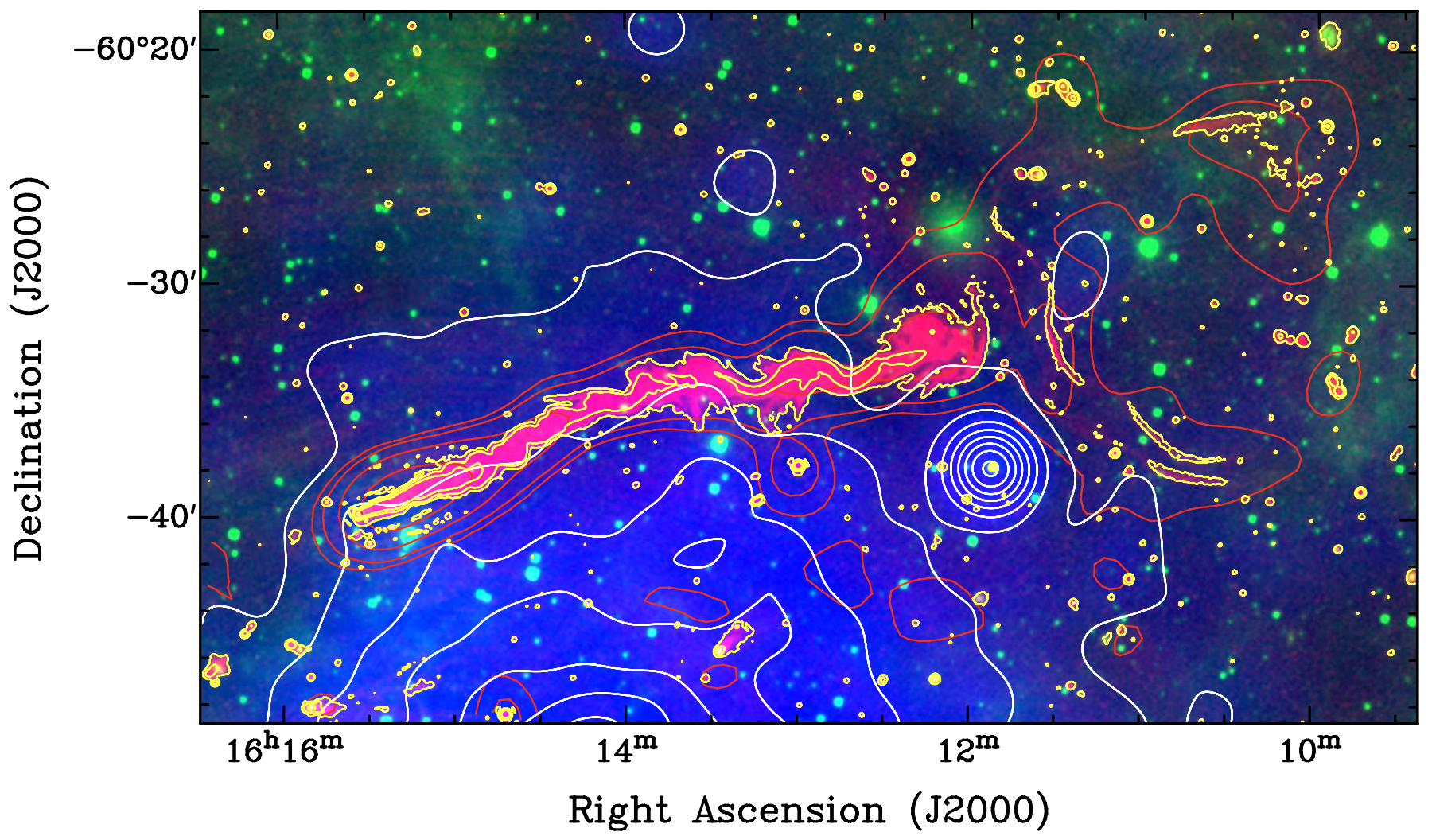}
\caption{Colour-composite image of the Corkscrew Galaxy in Abell~3627: ROSAT PSPC X-ray (blue + white contours at $\sim$20\% to 90\% of the peak flux, in steps of 10\%), ASKAP 943~MHz radio continuum (red + yellow contours at 0.2, 2, 20 and 200~mJy\,beam$^{-1}$) and WISE 12$\mu$m (green) images plus GLEAM 200~MHz contours in red at 0.1, 0.3, 1 and 3~mJy\,beam$^{-1}$. }
% ROSAT contours: 1, 1.5, 2, 2.5, 3, 3.5, 4, 4.5
% convolved to 150", then regridded
% R: -0.0001 to 0.0007 (EMU r=0)
% B: 0 to 2 (ROSAT PSPC 2deg)
% G: 390 to 425 (WISE 12 micron)
\label{fig:corkscrew-rgb}
\end{figure*}

Interactions between the turbulent ICM and the Corkscrew radio tail result in intricate filamentary structure in and beyond the $\sim$28\arcmin\ (570~kpc) long, helical radio tail, highlighted in Fig.~\ref{fig:Corkscrew-grey}. While the bright (young) part of its radio tail is highly collimated, the faint (old) part shows increasing oscillation amplitudes, break-ups, and filaments. The stunning set of arc-shaped radio filaments, discovered beyond and mostly orthogonal to the collimated Corkscrew tail end, forms a partial bubble (see Figs.~\ref{fig:corkscrew-emu+gleam} \& \ref{fig:corkscrew-rgb}). Face-on collision with a cold front in the cluster outskirts as simulated by \citet{Nolting2019} may have stripped the tail of its radio cocoon and created the bubble-shaped set of faint filaments -- the first detection of a "proto-lobe" -- which now forms the end of the Corkscrew tail, suggesting a full length of $\sim$45\arcmin\ or 920~kpc. Interactions of the radio galaxy tail with the ICM are likely responsible for the tail collimation and shear forces within the ICM for its increasingly filamentary structure. \\

Similar threads and magnetised filaments have been seen in  the neighbouring WAT 1610--60.8 \citep{Ramatsoku2020} and other cluster radio galaxies as well as, on much smaller scales, in the Galactic Centre \citep{Yusef-Zadeh1984,Heywood2019,Condon2021}. The latter are associated with the bipolar radio bubbles which form the vicinity of the central black hole. This can be explained by a simple model in which particles are streaming along pre-existing magnetic structures \citep[e.g.,][]{Vazza2021,Thomas2020,Rudnick2022}. For comparison, we consider the spectacular radio galaxy pair, NGC~1265 and NGC~1275, in the Perseus cluster \citep{Gendron-Marsolais2020}. The huge radio tail of NGC~1265 has complex filamentary structure, incl. a loop at the end of the bright tail and a long filament going SE after the first bend. Similar filamentary structure is also seen in other radio galaxy lobes and in diffuse cluster sources \citep{Rudnick2022,Condon2021,Botteon2024}. 

However, the much closer Corkscrew galaxy radio tail shows quite striking filaments sticking out at a wide range of angles, detailed in Fig.~\ref{fig:Corkscrew-grey}. Simulations of jets by, e.g., \citet{Chen2023} show filamentary structure, forward facing wisps, where transverse magnetic field pressure gradients can be balanced by ambient pressure gradients \citep[e.g.,][]{Porter2015}, see also \citet{ONeill2019b}. The long extent and helical shape of the Corkscrew Galaxy tail, which is located near the northern cluster periphery (see Fig.~\ref{fig:corkscrew-rgb}), together with the set of arc-shaped filaments west of its collimated tail end suggests a head-on collision with an orthogonal shock front as shown in 3D MHD simulations by \citet{Nolting2019}.

\subsection{Counter-jet} 

Almost all HT radio galaxies, when observed at sufficiently high resolution, show evidence for ejection of two jets from near their host galaxy's AGN \citep[e.g.,][]{deGregory2017}. So far, even in the highest resolution ASKAP images of the Corkscrew Galaxy, no twin tails are detected. But we do find the eastern-most radio emission peak offset from the optical galaxy (see Fig.~\ref{fig:corkscrew-bright}). The radio peak is at $\alpha,\delta$(J2000) = 16:15:33.3, --60:39:55.24, i.e. $3\arcsec \pm 1\arcsec$ ($\sim$1~kpc) east from the core of the bright elliptical host galaxy ESO\,137-G007 (2MASS J16153291--6039552). The offset was previously noted by \citet{JonesMcAdam1992}. It may be an indication of the missing counter-jet, which likely appears so short because it is bending backwards before merging in the wake of the host galaxy as discussed by \citet{JonesMcAdam1996}. This means that the western tail is actually a blend of both jets similar to the HT radio galaxy at the centre of Abell~2142 \citep{Venturi2017} which has been imaged at sub-arcsec resolution in \citet{deGregory2017}, and NGC~7385, a NAT radio galaxy with short, bent counter-tail \citep{Sebastian2017}.

\subsection{Polarisation}
The first polarisation maps of the Corkscrew Galaxy were made by \citet{JonesMcAdam1994,JonesMcAdam1996} using ATCA data at 1.36~GHz and 2.37~GHz (project C180). They reported the fractional polarisation to oscillate along the jet. \citet{JonesMcAdam1996} find the magnetic field to be parallel to the axis near the core and perpendicular in the bright part of the inner jet before getting too weak to measure. They also find the rotation measure (RM) to be large and highly variable. Our high-sensitivity ASKAP 944~MHz and 1.4~GHz maps allow for a detailed polarisation analysis along the full length of the radio tail, which will be presented in a follow-up paper (Anderson et al., in prep.).

\subsection{Helical corkscrew tail}

There is clear evidence for helical structures in the tail of the Corkscrew Galaxy, which have been noted before in a number of head-tail sources \citep[e.g.,][]{Sebastian2017} and in jets of normal double sources \citep[e.g.,][]{Asada2002, Zamaninasab2013} and at small scales in the M\,87 jet \citep{Pasetto2021}. The amplitude and clarity of the oscillations in the Corkscrew radio tail is greater than seen in any other source. We note that the amplitude of the corkscrew's sinusoidal oscillation near the end of the tail is $\sim$1\arcmin\ corresponding to 20~kpc, which is large compared to the helices seen previously in inner jets. The simulations by \citet[][their Fig.~5 at 492~Myr]{Nolting2019} show a similar "helical path" growing distance from the core, reminiscent of the Corkscrew Galaxy. \\

Similar to our radio galaxy pair, the HT galaxy IC\,711, which has a 900~kpc radio tail \citep{Srivastava2020}, is accompanied by the WAT radio galaxy IC\,708 spanning $\sim$80~kpc. \citet{Sebastian2017} provide an excellent review of both observations and models, including a good sample of HT radio galaxies with GMRT observations and multi-frequency spectral analysis. Two of their HT sources (NGC~1265 and PKS~B0053--016) clearly show the two inner jets bending to form one tail, and one HT source (NGC~7385) shows a a very short counter-tail that appears to bend backwards to form a joined tail. \citet{Sebastian2017} highlight the helical trajectory of the jets and symmetric wiggles in the tails of PKS~B0053--016, with the helicity likely due to the precession of the jets \citep[e.g.,][]{Blundell2004,Marscher2008}. They also show that the two jets seen in their high resolution images of HT galaxies appear as one jet/tail in their lower resolution images.

\section{Conclusions}

Using ASKAP wide-field images of the massive, merging galaxy cluster Abell~3627 ($D = 70$~Mpc) we focused on the head-tail radio galaxy 1610--60.5, better known as the Corkscrew Galaxy. We discovered intricate filamentary structures both along and beyond the Corkscrew Galaxy tail and summarise their properties and likely origin below. 

Re-processing of the calibrated ASKAP data from two observations at 944~MHz and 1.4~GHz, respectively, allowed us to make high-dynamic range radio continuum images at angular resolutions of $\sim$5\arcsec--15\arcsec\ as well as robust spectral index maps. This and the high surface-brightness sensitivity of ASKAP was essential for the discovery and analysis of these very faint features. \\

In the following we outline our main discoveries.

\begin{itemize}
   \item We find a stunning set of arc-shaped radio filaments / threads beyond and mostly orthogonal to the collimated Corkscrew tail, forming a partial bubble (diameter $\sim$ 15\arcmin) on the western periphery of the cluster. This may be the first detection of a "proto-lobe" similar to those seen in 3D MHD simulations by \citet{Nolting2019}, formed by the face-on impact of the Corkscrew Galaxy with a shock front in the cluster outskirts. The detected  synchrotron threads are very narrow ($\sim$20~kpc), long ($\sim$100~kpc), and faint with steep spectral indices. Including these filaments, the size of the Corkscrew galaxy tail spans $\sim$45\arcmin\ or 920~kpc.
   \item Furthermore, we find intricate filamentary structure along the older part of the Corkscrew galaxy tail, increasing in complexity towards the X-shaped tail end. While the bright part of the helical radio tail, emerging from near the central black hole of the galaxy ESO\,137-G007, is highly collimated \citep[see][]{JonesMcAdam1994,JonesMcAdam1996}, the older part shows increasing oscillation amplitudes, break-ups, and filaments. The latter consist of thin synchrotron threads (length $\sim$ 20 -- 80~kpc) pointing in various directions, likely shaped by the interactions of the Corkscrew tail with and shear forces in the surrounding ICM. We find a gradual steeping of the spectral index from the Corkscrew host galaxy to the X-shaped tail end, as expected by electron aging. 
  \item The helical structure of the Corkscrew Galaxy tail is one of its most well-known properties, persisting over its whole length ($\sim$28\arcmin\ or 570~kpc). We suggest that it is caused by precessing jets confined by a helical magnetic field. In the shock-normal scenario by \citet{Nolting2019} the aging radio tail is stripped of its outer cocoon and shows increased amplitude oscillations, having retained its original toroidal magnetic field.
  \item We find the brightest radio peak east of the host galaxy ($3\arcsec \pm 1\arcsec$ or $\sim$1~kpc offset), likely the start of a counter jet emerging from the near the black hole before bending backwards and becoming part of the Corkscrew Galaxy tail. Such offsets are also seen by \citet{Nolting2019} in their simulations of HT jets colliding with orthogonal shock fronts.
\end{itemize}

Further results are listed below. 
\begin{itemize}
  \item We confirm the numerous synchrotron threads within and between the lobes of the neighbouring WAT radio galaxy 1610--60.8, recently revealed by MeerKAT \citep{Ramatsoku2020}. 
  \item We find one new jellyfish galaxy (ESO\,137-G010) in the Abell~3627 cluster and detect radio tails (length $\sim$20--30~kpc) from this galaxy and the three already known jellyfish galaxies (ESO\,137-G001 and ESO\,137-G002, and WKK~6489), likely evidence of ram pressure stripping by the ICM. 
  \item We detect radio emission from the X-ray bright Seyfert\,1 galaxy WKK~6092, also known as IGR~J16119--6036 \citep{Nishino2012}.  
\end{itemize}

\section*{Acknowledgments}

We thank Bi-Qing For and Ron Ekers for comments on an early version of this paper. AB acknowledges financial support from the European Union -- Next Generation EU. -- This scientific work uses data obtained from Inyarrimanha Ilgari Bundara / the Murchison Radio-astronomy Observatory. We acknowledge the Wajarri Yamaji People as the Traditional Owners and native title holders of the Observatory site. CSIRO’s ASKAP radio telescope is part of the Australia Telescope National Facility (\url{https://ror.org/05qajvd42}). Operation of ASKAP is funded by the Australian Government with support from the National Collaborative Research Infrastructure Strategy. ASKAP uses the resources of the Pawsey Supercomputing Research Centre. Establishment of ASKAP, Inyarrimanha Ilgari Bundara, the CSIRO Murchison Radio-astronomy Observatory and the Pawsey Supercomputing Research Centre are initiatives of the Australian Government, with support from the Government of Western Australia and the Science and Industry Endowment Fund. This paper includes archived data obtained through the CSIRO ASKAP Science Data Archive, CASDA (\url{http://data.csiro.au}).

\section*{Data availability}
ASKAP data products are publicly available in CASDA.
% the CSIRO ASKAP Science Data Archive (CASDA) at \texttt{data.csiro.au/domain/casdaObservation}. The re-processed images will also be made available via CASDA at doi-tbd.

% The Acknowledgements section is not numbered. Here you can thank helpful colleagues, acknowledge funding agencies, telescopes and facilities used etc. Try to keep it short.

% \clearpage
% \newpage
% \appendix

%%%%%%%%%%%%%%%%%%%%%%%%%%%%%%%%%%%%%%%%%%%%%%%%%%



\bibliographystyle{mnras}
\bibliography{A3627}

\begin{thebibliography}{}
\makeatletter
\relax
\def\mn@urlcharsother{\let\do\@makeother \do\$\do\&\do\#\do\^\do\_\do\%\do\~}
\def\mn@doi{\begingroup\mn@urlcharsother \@ifnextchar [ {\mn@doi@} {\mn@doi@[]}}
\def\mn@doi@[#1]#2{\def\@tempa{#1}\ifx\@tempa\@empty \href {http://dx.doi.org/#2} {doi:#2}\else \href {http://dx.doi.org/#2} {#1}\fi \endgroup}
\def\mn@eprint#1#2{\mn@eprint@#1:#2::\@nil}
\def\mn@eprint@arXiv#1{\href {http://arxiv.org/abs/#1} {{\tt arXiv:#1}}}
\def\mn@eprint@dblp#1{\href {http://dblp.uni-trier.de/rec/bibtex/#1.xml} {dblp:#1}}
\def\mn@eprint@#1:#2:#3:#4\@nil{\def\@tempa {#1}\def\@tempb {#2}\def\@tempc {#3}\ifx \@tempc \@empty \let \@tempc \@tempb \let \@tempb \@tempa \fi \ifx \@tempb \@empty \def\@tempb {arXiv}\fi \@ifundefined {mn@eprint@\@tempb}{\@tempb:\@tempc}{\expandafter \expandafter \csname mn@eprint@\@tempb\endcsname \expandafter{\@tempc}}}

\bibitem[\protect\citeauthoryear{{Abell}, {Corwin}  \& {Olowin}}{{Abell} et~al.}{1989}]{Abell1989}
{Abell} G.~O.,  {Corwin} Harold~G. J.,   {Olowin} R.~P.,  1989, \mn@doi [\apjs] {10.1086/191333}, \href {https://ui.adsabs.harvard.edu/abs/1989ApJS...70....1A} {70, 1}

\bibitem[\protect\citeauthoryear{{Asada}, {Inoue}, {Uchida}, {Kameno}, {Fujisawa}, {Iguchi}  \& {Mutoh}}{{Asada} et~al.}{2002}]{Asada2002}
{Asada} K.,  {Inoue} M.,  {Uchida} Y.,  {Kameno} S.,  {Fujisawa} K.,  {Iguchi} S.,   {Mutoh} M.,  2002, \mn@doi [\pasj] {10.1093/pasj/54.3.L39}, \href {https://ui.adsabs.harvard.edu/abs/2002PASJ...54L..39A} {54, L39}

\bibitem[\protect\citeauthoryear{{Blundell} \& {Bowler}}{{Blundell} \& {Bowler}}{2004}]{Blundell2004}
{Blundell} K.~M.,  {Bowler} M.~G.,  2004, \mn@doi [\apjl] {10.1086/426542}, \href {https://ui.adsabs.harvard.edu/abs/2004ApJ...616L.159B} {616, L159}

\bibitem[\protect\citeauthoryear{{Boehringer}, {Neumann}, {Schindler}  \& {Kraan-Korteweg}}{{Boehringer} et~al.}{1996}]{Boehringer1996}
{Boehringer} H.,  {Neumann} D.~M.,  {Schindler} S.,   {Kraan-Korteweg} R.~C.,  1996, \mn@doi [\apj] {10.1086/177592}, \href {https://ui.adsabs.harvard.edu/abs/1996ApJ...467..168B} {467, 168}

\bibitem[\protect\citeauthoryear{{Botteon} et~al.,}{{Botteon} et~al.}{2020}]{Botteon2020}
{Botteon} A.,  et~al., 2020, \mn@doi [\apj] {10.3847/1538-4357/ab9a2f}, \href {https://ui.adsabs.harvard.edu/abs/2020ApJ...897...93B} {897, 93}

\bibitem[\protect\citeauthoryear{{Botteon} et~al.,}{{Botteon} et~al.}{2024}]{Botteon2024}
{Botteon} A.,  et~al., 2024, \mn@doi [\mnras] {10.1093/mnras/stad3305}, \href {https://ui.adsabs.harvard.edu/abs/2024MNRAS.527..919B} {527, 919}

\bibitem[\protect\citeauthoryear{{Briggs}}{{Briggs}}{1995}]{Briggs95}
{Briggs} D.~S.,  1995, PhD thesis, The New Mexico Institute of Mining and Technology, Socorro, New Mexico

\bibitem[\protect\citeauthoryear{{Bruno} et~al.,}{{Bruno} et~al.}{2023}]{Bruno2023}
{Bruno} L.,  et~al., 2023, \mn@doi [\aap] {10.1051/0004-6361/202347245}, \href {https://ui.adsabs.harvard.edu/abs/2023A&A...678A.133B} {678, A133}

\bibitem[\protect\citeauthoryear{{CASA Team} et~al.,}{{CASA Team} et~al.}{2022}]{CASA2022}
{CASA Team} et~al., 2022, \mn@doi [\pasp] {10.1088/1538-3873/ac9642}, \href {https://ui.adsabs.harvard.edu/abs/2022PASP..134k4501C} {134, 114501}

\bibitem[\protect\citeauthoryear{{Chen}, {Heinz}  \& {Hooper}}{{Chen} et~al.}{2023}]{Chen2023}
{Chen} Y.-H.,  {Heinz} S.,   {Hooper} E.,  2023, \mn@doi [\mnras] {10.1093/mnras/stad1074}, \href {https://ui.adsabs.harvard.edu/abs/2023MNRAS.522.2850C} {522, 2850}

\bibitem[\protect\citeauthoryear{{Chippendale} et~al.,}{{Chippendale} et~al.}{2015}]{Chippendale2015}
{Chippendale} A.~P.,  et~al., 2015, in 2015 International Conference on Electromagnetics in Advanced Applications (ICEAA. pp 541--544 (\mn@eprint {arXiv} {1509.00544}), \mn@doi{10.1109/ICEAA.2015.7297174}

\bibitem[\protect\citeauthoryear{{Christiansen}, {Frater}, {Watkinson}, {O'Sullivan}, {Lockhart}  \& {Goss}}{{Christiansen} et~al.}{1977}]{Christiansen1977}
{Christiansen} W.~N.,  {Frater} R.~H.,  {Watkinson} A.,  {O'Sullivan} J.~D.,  {Lockhart} I.~A.,   {Goss} W.~M.,  1977, \mn@doi [\mnras] {10.1093/mnras/181.2.183}, \href {https://ui.adsabs.harvard.edu/abs/1977MNRAS.181..183C} {181, 183}

\bibitem[\protect\citeauthoryear{{Condon}, {Cotton}, {White}, {Legodi}, {Goedhart}, {McAlpine}, {Ratcliffe}  \& {Camilo}}{{Condon} et~al.}{2021}]{Condon2021}
{Condon} J.~J.,  {Cotton} W.~D.,  {White} S.~V.,  {Legodi} S.,  {Goedhart} S.,  {McAlpine} K.,  {Ratcliffe} S.~M.,   {Camilo} F.,  2021, \mn@doi [\apj] {10.3847/1538-4357/ac0880}, \href {https://ui.adsabs.harvard.edu/abs/2021ApJ...917...18C} {917, 18}

\bibitem[\protect\citeauthoryear{{Cuciti}, {Brunetti}, {van Weeren}, {Bonafede}, {Dallacasa}, {Cassano}, {Venturi}  \& {Kale}}{{Cuciti} et~al.}{2018}]{Cuciti2018}
{Cuciti} V.,  {Brunetti} G.,  {van Weeren} R.,  {Bonafede} A.,  {Dallacasa} D.,  {Cassano} R.,  {Venturi} T.,   {Kale} R.,  2018, \mn@doi [\aap] {10.1051/0004-6361/201731174}, \href {https://ui.adsabs.harvard.edu/abs/2018A&A...609A..61C} {609, A61}

\bibitem[\protect\citeauthoryear{{Dressler}, {Faber}, {Burstein}, {Davies}, {Lynden-Bell}, {Terlevich}  \& {Wegner}}{{Dressler} et~al.}{1987}]{Dressler1987}
{Dressler} A.,  {Faber} S.~M.,  {Burstein} D.,  {Davies} R.~L.,  {Lynden-Bell} D.,  {Terlevich} R.~J.,   {Wegner} G.,  1987, \mn@doi [\apjl] {10.1086/184827}, \href {https://ui.adsabs.harvard.edu/abs/1987ApJ...313L..37D} {313, L37}

\bibitem[\protect\citeauthoryear{{Duchesne} et~al.,}{{Duchesne} et~al.}{2023}]{Duchesne2023}
{Duchesne} S.~W.,  et~al., 2023, \mn@doi [\pasa] {10.1017/pasa.2023.31}, \href {https://ui.adsabs.harvard.edu/abs/2023PASA...40...34D} {40, e034}

\bibitem[\protect\citeauthoryear{{Duchesne} et~al.,}{{Duchesne} et~al.}{2024}]{Duchesne2024}
{Duchesne} S.~W.,  et~al., 2024, arXiv e-prints, \href {https://ui.adsabs.harvard.edu/abs/2024arXiv240206192D} {p. arXiv:2402.06192}

\bibitem[\protect\citeauthoryear{{Ekers}}{{Ekers}}{1969}]{Ekers1969}
{Ekers} R.~D.,  1969, Australian Journal of Physics Astrophysical Supplement, \href {https://ui.adsabs.harvard.edu/abs/1969AuJPA...6....3E} {6, 3}

\bibitem[\protect\citeauthoryear{{Ekers}}{{Ekers}}{1970}]{Ekers1970}
{Ekers} R.~D.,  1970, \mn@doi [Australian Journal of Physics] {10.1071/PH700217}, \href {https://ui.adsabs.harvard.edu/abs/1970AuJPh..23..217E} {23, 217}

\bibitem[\protect\citeauthoryear{{Forman} \& {Jones}}{{Forman} \& {Jones}}{1982}]{Forman1982}
{Forman} W.,  {Jones} C.,  1982, \mn@doi [\araa] {10.1146/annurev.aa.20.090182.002555}, \href {https://ui.adsabs.harvard.edu/abs/1982ARA&A..20..547F} {20, 547}

\bibitem[\protect\citeauthoryear{{Fumagalli}, {Fossati}, {Hau}, {Gavazzi}, {Bower}, {Sun}  \& {Boselli}}{{Fumagalli} et~al.}{2014}]{Fumagalli2014}
{Fumagalli} M.,  {Fossati} M.,  {Hau} G. K.~T.,  {Gavazzi} G.,  {Bower} R.,  {Sun} M.,   {Boselli} A.,  2014, \mn@doi [\mnras] {10.1093/mnras/stu2092}, \href {https://ui.adsabs.harvard.edu/abs/2014MNRAS.445.4335F} {445, 4335}

\bibitem[\protect\citeauthoryear{{Gendron-Marsolais} et~al.,}{{Gendron-Marsolais} et~al.}{2020}]{Gendron-Marsolais2020}
{Gendron-Marsolais} M.,  et~al., 2020, \mn@doi [\mnras] {10.1093/mnras/staa2003}, \href {https://ui.adsabs.harvard.edu/abs/2020MNRAS.499.5791G} {499, 5791}

\bibitem[\protect\citeauthoryear{{Hancock}, {Murphy}, {Gaensler}, {Hopkins}  \& {Curran}}{{Hancock} et~al.}{2012}]{aegean1}
{Hancock} P.~J.,  {Murphy} T.,  {Gaensler} B.~M.,  {Hopkins} A.,   {Curran} J.~R.,  2012, \mn@doi [\mnras] {10.1111/j.1365-2966.2012.20768.x}, \href {https://ui.adsabs.harvard.edu/abs/2012MNRAS.422.1812H} {422, 1812}

\bibitem[\protect\citeauthoryear{{Hancock}, {Trott}  \& {Hurley-Walker}}{{Hancock} et~al.}{2018}]{aegean2}
{Hancock} P.~J.,  {Trott} C.~M.,   {Hurley-Walker} N.,  2018, \mn@doi [\pasa] {10.1017/pasa.2018.3}, \href {https://ui.adsabs.harvard.edu/abs/2018PASA...35...11H} {35, e011}

\bibitem[\protect\citeauthoryear{{Heywood} et~al.,}{{Heywood} et~al.}{2019}]{Heywood2019}
{Heywood} I.,  et~al., 2019, \mn@doi [\nat] {10.1038/s41586-019-1532-5}, \href {https://ui.adsabs.harvard.edu/abs/2019Natur.573..235H} {573, 235}

\bibitem[\protect\citeauthoryear{{Hotan} et~al.,}{{Hotan} et~al.}{2021}]{Hotan2021}
{Hotan} A.~W.,  et~al., 2021, \mn@doi [\pasa] {10.1017/pasa.2021.1}, \href {https://ui.adsabs.harvard.edu/abs/2021PASA...38....9H} {38, e009}

\bibitem[\protect\citeauthoryear{{Hurley-Walker} et~al.,}{{Hurley-Walker} et~al.}{2017}]{Hurley-Walker2017}
{Hurley-Walker} N.,  et~al., 2017, \mn@doi [\mnras] {10.1093/mnras/stw2337}, \href {https://ui.adsabs.harvard.edu/abs/2017MNRAS.464.1146H} {464, 1146}

\bibitem[\protect\citeauthoryear{{Ignesti} et~al.,}{{Ignesti} et~al.}{2023}]{Ignesti2023}
{Ignesti} A.,  et~al., 2023, \mn@doi [\aap] {10.1051/0004-6361/202346517}, \href {https://ui.adsabs.harvard.edu/abs/2023A&A...675A.118I} {675, A118}

\bibitem[\protect\citeauthoryear{{Johnston} et~al.,}{{Johnston} et~al.}{2008}]{Johnston2008}
{Johnston} S.,  et~al., 2008, \mn@doi [Experimental Astronomy] {10.1007/s10686-008-9124-7}, \href {https://ui.adsabs.harvard.edu/abs/2008ExA....22..151J} {22, 151}

\bibitem[\protect\citeauthoryear{{Jones} \& {McAdam}}{{Jones} \& {McAdam}}{1992}]{JonesMcAdam1992}
{Jones} P.~A.,  {McAdam} W.~B.,  1992, \mn@doi [\apjs] {10.1086/191662}, \href {https://ui.adsabs.harvard.edu/abs/1992ApJS...80..137J} {80, 137}

\bibitem[\protect\citeauthoryear{{Jones} \& {McAdam}}{{Jones} \& {McAdam}}{1994}]{JonesMcAdam1994}
{Jones} P.~A.,  {McAdam} W.~B.,  1994, \mn@doi [\pasa] {10.1017/S132335800001972X}, \href {https://ui.adsabs.harvard.edu/abs/1994PASA...11...74J} {11, 74}

\bibitem[\protect\citeauthoryear{{Jones} \& {McAdam}}{{Jones} \& {McAdam}}{1996}]{JonesMcAdam1996}
{Jones} P.~A.,  {McAdam} W.~B.,  1996, \mn@doi [\mnras] {10.1093/mnras/282.1.137}, \href {https://ui.adsabs.harvard.edu/abs/1996MNRAS.282..137J} {282, 137}

\bibitem[\protect\citeauthoryear{{Koribalski}}{{Koribalski}}{2012}]{Koribalski2012}
{Koribalski} B.~S.,  2012, \mn@doi [\pasa] {10.1071/AS12030}, \href {https://ui.adsabs.harvard.edu/abs/2012PASA...29..359K} {29, 359}

\bibitem[\protect\citeauthoryear{Koribalski}{Koribalski}{2022}]{Koribalski2022}
Koribalski B.~S.,  2022, in 2022 3rd URSI Atlantic and Asia Pacific Radio Science Meeting (AT-AP-RASC). pp~1--4, \mn@doi{10.23919/AT-AP-RASC54737.2022.9814179}

\bibitem[\protect\citeauthoryear{{Koribalski} et~al.,}{{Koribalski} et~al.}{2020}]{Koribalski2020}
{Koribalski} B.~S.,  et~al., 2020, \mn@doi [\apss] {10.1007/s10509-020-03831-4}, \href {https://ui.adsabs.harvard.edu/abs/2020Ap&SS.365..118K} {365, 118}

\bibitem[\protect\citeauthoryear{{Kraan-Korteweg}, {Woudt}, {Cayatte}, {Fairall}, {Balkowski}  \& {Henning}}{{Kraan-Korteweg} et~al.}{1996}]{Kraan1996}
{Kraan-Korteweg} R.~C.,  {Woudt} P.~A.,  {Cayatte} V.,  {Fairall} A.~P.,  {Balkowski} C.,   {Henning} P.~A.,  1996, \mn@doi [\nat] {10.1038/379519a0}, \href {https://ui.adsabs.harvard.edu/abs/1996Natur.379..519K} {379, 519}

\bibitem[\protect\citeauthoryear{{Lal} et~al.,}{{Lal} et~al.}{2022}]{Lal2022}
{Lal} D.~V.,  et~al., 2022, \mn@doi [\apj] {10.3847/1538-4357/ac7a9b}, \href {https://ui.adsabs.harvard.edu/abs/2022ApJ...934..170L} {934, 170}

\bibitem[\protect\citeauthoryear{{Lauberts} \& {Valentijn}}{{Lauberts} \& {Valentijn}}{1989}]{ESO-LV1989}
{Lauberts} A.,  {Valentijn} E.~A.,  1989, {The surface photometry catalogue of the ESO-Uppsala galaxies}

\bibitem[\protect\citeauthoryear{{Laudari} et~al.,}{{Laudari} et~al.}{2022}]{Laudari2022}
{Laudari} S.,  et~al., 2022, \mn@doi [\mnras] {10.1093/mnras/stab3280}, \href {https://ui.adsabs.harvard.edu/abs/2022MNRAS.509.3938L} {509, 3938}

\bibitem[\protect\citeauthoryear{{Lynden-Bell}, {Faber}, {Burstein}, {Davies}, {Dressler}, {Terlevich}  \& {Wegner}}{{Lynden-Bell} et~al.}{1988}]{Lynden-Bell1988}
{Lynden-Bell} D.,  {Faber} S.~M.,  {Burstein} D.,  {Davies} R.~L.,  {Dressler} A.,  {Terlevich} R.~J.,   {Wegner} G.,  1988, \mn@doi [\apj] {10.1086/166066}, \href {https://ui.adsabs.harvard.edu/abs/1988ApJ...326...19L} {326, 19}

\bibitem[\protect\citeauthoryear{{Marscher} et~al.,}{{Marscher} et~al.}{2008}]{Marscher2008}
{Marscher} A.~P.,  et~al., 2008, \mn@doi [\nat] {10.1038/nature06895}, \href {https://ui.adsabs.harvard.edu/abs/2008Natur.452..966M} {452, 966}

\bibitem[\protect\citeauthoryear{{McConnell} et~al.,}{{McConnell} et~al.}{2020}]{McConnell2020}
{McConnell} D.,  et~al., 2020, \mn@doi [\pasa] {10.1017/pasa.2020.41}, \href {https://ui.adsabs.harvard.edu/abs/2020PASA...37...48M} {37, e048}

\bibitem[\protect\citeauthoryear{{Morris}, {Wilcots}, {Hooper}  \& {Heinz}}{{Morris} et~al.}{2022}]{Morris2022}
{Morris} M.~E.,  {Wilcots} E.,  {Hooper} E.,   {Heinz} S.,  2022, \mn@doi [\aj] {10.3847/1538-3881/ac66db}, \href {https://ui.adsabs.harvard.edu/abs/2022AJ....163..280M} {163, 280}

\bibitem[\protect\citeauthoryear{{M{\"u}ller} et~al.,}{{M{\"u}ller} et~al.}{2021}]{Mueller2021}
{M{\"u}ller} A.,  et~al., 2021, \mn@doi [\mnras] {10.1093/mnras/stab2928}, \href {https://ui.adsabs.harvard.edu/abs/2021MNRAS.508.5326M} {508, 5326}

\bibitem[\protect\citeauthoryear{{Nishino}, {Fukazawa}  \& {Hayashi}}{{Nishino} et~al.}{2012}]{Nishino2012}
{Nishino} S.,  {Fukazawa} Y.,   {Hayashi} K.,  2012, \mn@doi [\pasj] {10.1093/pasj/64.1.16}, \href {https://ui.adsabs.harvard.edu/abs/2012PASJ...64...16N} {64, 16}

\bibitem[\protect\citeauthoryear{{Nolting}, {Jones}, {O'Neill}  \& {Mendygral}}{{Nolting} et~al.}{2019a}]{Nolting2019}
{Nolting} C.,  {Jones} T.~W.,  {O'Neill} B.~J.,   {Mendygral} P.~J.,  2019a, \mn@doi [\apj] {10.3847/1538-4357/ab16d6}, \href {https://ui.adsabs.harvard.edu/abs/2019ApJ...876..154N} {876, 154}

\bibitem[\protect\citeauthoryear{{Nolting}, {Jones}, {O'Neill}  \& {Mendygral}}{{Nolting} et~al.}{2019b}]{Nolting-II2019}
{Nolting} C.,  {Jones} T.~W.,  {O'Neill} B.~J.,   {Mendygral} P.~J.,  2019b, \mn@doi [\apj] {10.3847/1538-4357/ab4650}, \href {https://ui.adsabs.harvard.edu/abs/2019ApJ...885...80N} {885, 80}

\bibitem[\protect\citeauthoryear{{Norris} et~al.,}{{Norris} et~al.}{2011}]{EMU}
{Norris} R.~P.,  et~al., 2011, \mn@doi [\pasa] {10.1071/AS11021}, \href {https://ui.adsabs.harvard.edu/abs/2011PASA...28..215N} {28, 215}

\bibitem[\protect\citeauthoryear{{Norris} et~al.,}{{Norris} et~al.}{2021}]{EMU-PS}
{Norris} R.~P.,  et~al., 2021, \mn@doi [\pasa] {10.1017/pasa.2021.42}, \href {https://ui.adsabs.harvard.edu/abs/2021PASA...38...46N} {38, e046}

\bibitem[\protect\citeauthoryear{{O'Dea}, {Sarazin}  \& {Owen}}{{O'Dea} et~al.}{1987}]{ODea1987}
{O'Dea} C.~P.,  {Sarazin} C.~L.,   {Owen} F.~N.,  1987, \mn@doi [\apj] {10.1086/165183}, \href {https://ui.adsabs.harvard.edu/abs/1987ApJ...316..113O} {316, 113}

\bibitem[\protect\citeauthoryear{{O'Neill}, {Jones}, {Nolting}  \& {Mendygral}}{{O'Neill} et~al.}{2019}]{ONeill2019b}
{O'Neill} B.~J.,  {Jones} T.~W.,  {Nolting} C.,   {Mendygral} P.~J.,  2019, \mn@doi [\apj] {10.3847/1538-4357/ab4efa}, \href {https://ui.adsabs.harvard.edu/abs/2019ApJ...887...26O} {887, 26}

\bibitem[\protect\citeauthoryear{{Offringa} \& {Smirnov}}{{Offringa} \& {Smirnov}}{2017}]{Offringa2017}
{Offringa} A.~R.,  {Smirnov} O.,  2017, \mn@doi [\mnras] {10.1093/mnras/stx1547}, \href {https://ui.adsabs.harvard.edu/abs/2017MNRAS.471..301O} {471, 301}

\bibitem[\protect\citeauthoryear{{Offringa} et~al.,}{{Offringa} et~al.}{2014}]{Offringa2014}
{Offringa} A.~R.,  et~al., 2014, \mn@doi [\mnras] {10.1093/mnras/stu1368}, \href {https://ui.adsabs.harvard.edu/abs/2014MNRAS.444..606O} {444, 606}

\bibitem[\protect\citeauthoryear{{Owen}, {Rudnick}, {Eilek}, {Rau}, {Bhatnagar}  \& {Kogan}}{{Owen} et~al.}{2014}]{Owen2014}
{Owen} F.~N.,  {Rudnick} L.,  {Eilek} J.,  {Rau} U.,  {Bhatnagar} S.,   {Kogan} L.,  2014, \mn@doi [\apj] {10.1088/0004-637X/794/1/24}, \href {https://ui.adsabs.harvard.edu/abs/2014ApJ...794...24O} {794, 24}

\bibitem[\protect\citeauthoryear{{Pasetto} et~al.,}{{Pasetto} et~al.}{2021}]{Pasetto2021}
{Pasetto} A.,  et~al., 2021, \mn@doi [\apjl] {10.3847/2041-8213/ac3a88}, \href {https://ui.adsabs.harvard.edu/abs/2021ApJ...923L...5P} {923, L5}

\bibitem[\protect\citeauthoryear{{Planck Collaboration} et~al.,}{{Planck Collaboration} et~al.}{2016}]{Planck2016}
{Planck Collaboration} et~al., 2016, \mn@doi [\aap] {10.1051/0004-6361/201525823}, \href {https://ui.adsabs.harvard.edu/abs/2016A&A...594A..27P} {594, A27}

\bibitem[\protect\citeauthoryear{{Porter}, {Jones}  \& {Ryu}}{{Porter} et~al.}{2015}]{Porter2015}
{Porter} D.~H.,  {Jones} T.~W.,   {Ryu} D.,  2015, \mn@doi [\apj] {10.1088/0004-637X/810/2/93}, \href {https://ui.adsabs.harvard.edu/abs/2015ApJ...810...93P} {810, 93}

\bibitem[\protect\citeauthoryear{{Ramatsoku} et~al.,}{{Ramatsoku} et~al.}{2020}]{Ramatsoku2020}
{Ramatsoku} M.,  et~al., 2020, \mn@doi [\aap] {10.1051/0004-6361/202037800}, \href {https://ui.adsabs.harvard.edu/abs/2020A&A...636L...1R} {636, L1}

\bibitem[\protect\citeauthoryear{{Riseley} et~al.,}{{Riseley} et~al.}{2022}]{Riseley2022}
{Riseley} C.~J.,  et~al., 2022, \mn@doi [\mnras] {10.1093/mnras/stac1771}, \href {https://ui.adsabs.harvard.edu/abs/2022MNRAS.515.1871R} {515, 1871}

\bibitem[\protect\citeauthoryear{{Roberts}, {van Weeren}, {Timmerman}, {Botteon}, {Gendron-Marsolais}, {Ignesti}  \& {Rottgering}}{{Roberts} et~al.}{2022}]{Roberts2022}
{Roberts} I.~D.,  {van Weeren} R.~J.,  {Timmerman} R.,  {Botteon} A.,  {Gendron-Marsolais} M.,  {Ignesti} A.,   {Rottgering} H.~J.~A.,  2022, \mn@doi [\aap] {10.1051/0004-6361/202142294}, \href {https://ui.adsabs.harvard.edu/abs/2022A&A...658A..44R} {658, A44}

\bibitem[\protect\citeauthoryear{{Rottgering}, {Snellen}, {Miley}, {de Jong}, {Hanisch}  \& {Perley}}{{Rottgering} et~al.}{1994}]{Rottgering1994}
{Rottgering} H.,  {Snellen} I.,  {Miley} G.,  {de Jong} J.~P.,  {Hanisch} R.~J.,   {Perley} R.,  1994, \mn@doi [\apj] {10.1086/174940}, \href {https://ui.adsabs.harvard.edu/abs/1994ApJ...436..654R} {436, 654}

\bibitem[\protect\citeauthoryear{{Rudnick} et~al.,}{{Rudnick} et~al.}{2022}]{Rudnick2022}
{Rudnick} L.,  et~al., 2022, \mn@doi [\apj] {10.3847/1538-4357/ac7c76}, \href {https://ui.adsabs.harvard.edu/abs/2022ApJ...935..168R} {935, 168}

\bibitem[\protect\citeauthoryear{{Sarazin}}{{Sarazin}}{1986}]{Sarazin1986}
{Sarazin} C.~L.,  1986, \mn@doi [Reviews of Modern Physics] {10.1103/RevModPhys.58.1}, \href {https://ui.adsabs.harvard.edu/abs/1986RvMP...58....1S} {58, 1}

\bibitem[\protect\citeauthoryear{{Sebastian}, {Lal}  \& {Pramesh Rao}}{{Sebastian} et~al.}{2017}]{Sebastian2017}
{Sebastian} B.,  {Lal} D.~V.,   {Pramesh Rao} A.,  2017, \mn@doi [\aj] {10.3847/1538-3881/aa88d0}, \href {https://ui.adsabs.harvard.edu/abs/2017AJ....154..169S} {154, 169}

\bibitem[\protect\citeauthoryear{{Skrutskie} et~al.,}{{Skrutskie} et~al.}{2006}]{Skrutskie2006}
{Skrutskie} M.~F.,  et~al., 2006, \mn@doi [\aj] {10.1086/498708}, \href {https://ui.adsabs.harvard.edu/abs/2006AJ....131.1163S} {131, 1163}

\bibitem[\protect\citeauthoryear{{Srivastava} \& {Singal}}{{Srivastava} \& {Singal}}{2020}]{Srivastava2020}
{Srivastava} S.,  {Singal} A.~K.,  2020, \mn@doi [\mnras] {10.1093/mnras/staa520}, \href {https://ui.adsabs.harvard.edu/abs/2020MNRAS.493.3811S} {493, 3811}

\bibitem[\protect\citeauthoryear{{Sun}, {Jones}, {Forman}, {Nulsen}, {Donahue}  \& {Voit}}{{Sun} et~al.}{2006}]{Sun2006}
{Sun} M.,  {Jones} C.,  {Forman} W.,  {Nulsen} P.~E.~J.,  {Donahue} M.,   {Voit} G.~M.,  2006, \mn@doi [\apjl] {10.1086/500590}, \href {https://ui.adsabs.harvard.edu/abs/2006ApJ...637L..81S} {637, L81}

\bibitem[\protect\citeauthoryear{{Sun}, {Donahue}  \& {Voit}}{{Sun} et~al.}{2007}]{Sun2007}
{Sun} M.,  {Donahue} M.,   {Voit} G.~M.,  2007, \mn@doi [\apj] {10.1086/522690}, \href {https://ui.adsabs.harvard.edu/abs/2007ApJ...671..190S} {671, 190}

\bibitem[\protect\citeauthoryear{{Sun}, {Donahue}, {Roediger}, {Nulsen}, {Voit}, {Sarazin}, {Forman}  \& {Jones}}{{Sun} et~al.}{2010}]{Sun2010}
{Sun} M.,  {Donahue} M.,  {Roediger} E.,  {Nulsen} P.~E.~J.,  {Voit} G.~M.,  {Sarazin} C.,  {Forman} W.,   {Jones} C.,  2010, \mn@doi [\apj] {10.1088/0004-637X/708/2/946}, \href {https://ui.adsabs.harvard.edu/abs/2010ApJ...708..946S} {708, 946}

\bibitem[\protect\citeauthoryear{{Tamura}, {Fukazawa}, {Kaneda}, {Makishima}, {Tashiro}, {Tanaka}  \& {Bohringer}}{{Tamura} et~al.}{1998}]{Tamura1998}
{Tamura} T.,  {Fukazawa} Y.,  {Kaneda} H.,  {Makishima} K.,  {Tashiro} M.,  {Tanaka} Y.,   {Bohringer} H.,  1998, \mn@doi [\pasj] {10.1093/pasj/50.2.195}, \href {https://ui.adsabs.harvard.edu/abs/1998PASJ...50..195T} {50, 195}

\bibitem[\protect\citeauthoryear{{Terni de Gregory}, {Feretti}, {Giovannini}, {Govoni}, {Murgia}, {Perley}  \& {Vacca}}{{Terni de Gregory} et~al.}{2017}]{deGregory2017}
{Terni de Gregory} B.,  {Feretti} L.,  {Giovannini} G.,  {Govoni} F.,  {Murgia} M.,  {Perley} R.~A.,   {Vacca} V.,  2017, \mn@doi [\aap] {10.1051/0004-6361/201730878}, \href {https://ui.adsabs.harvard.edu/abs/2017A&A...608A..58T} {608, A58}

\bibitem[\protect\citeauthoryear{{Thomas}, {Pfrommer}  \& {En{\ss}lin}}{{Thomas} et~al.}{2020}]{Thomas2020}
{Thomas} T.,  {Pfrommer} C.,   {En{\ss}lin} T.,  2020, \mn@doi [\apjl] {10.3847/2041-8213/ab7237}, \href {https://ui.adsabs.harvard.edu/abs/2020ApJ...890L..18T} {890, L18}

\bibitem[\protect\citeauthoryear{{Vazza} \& {Botteon}}{{Vazza} \& {Botteon}}{2024}]{VazzaBotteon2024}
{Vazza} F.,  {Botteon} A.,  2024, \mn@doi [arXiv e-prints] {10.48550/arXiv.2403.16068}, \href {https://ui.adsabs.harvard.edu/abs/2024arXiv240316068V} {p. arXiv:2403.16068}

\bibitem[\protect\citeauthoryear{{Vazza}, {Wittor}, {Brunetti}  \& {Br{\"u}ggen}}{{Vazza} et~al.}{2021}]{Vazza2021}
{Vazza} F.,  {Wittor} D.,  {Brunetti} G.,   {Br{\"u}ggen} M.,  2021, \mn@doi [\aap] {10.1051/0004-6361/202140513}, \href {https://ui.adsabs.harvard.edu/abs/2021A&A...653A..23V} {653, A23}

\bibitem[\protect\citeauthoryear{{Vazza}, {Wittor}, {Di Federico}, {Br{\"u}ggen}, {Brienza}, {Brunetti}, {Brighenti}  \& {Pasini}}{{Vazza} et~al.}{2023}]{Vazza2023}
{Vazza} F.,  {Wittor} D.,  {Di Federico} L.,  {Br{\"u}ggen} M.,  {Brienza} M.,  {Brunetti} G.,  {Brighenti} F.,   {Pasini} T.,  2023, \mn@doi [\aap] {10.1051/0004-6361/202243753}, \href {https://ui.adsabs.harvard.edu/abs/2023A&A...669A..50V} {669, A50}

\bibitem[\protect\citeauthoryear{{Velovi{\'c}} et~al.,}{{Velovi{\'c}} et~al.}{2022}]{Velovic2022}
{Velovi{\'c}} V.,  et~al., 2022, \mn@doi [\mnras] {10.1093/mnras/stac2012}, \href {https://ui.adsabs.harvard.edu/abs/2022MNRAS.516.1865V} {516, 1865}

\bibitem[\protect\citeauthoryear{{Venturi} et~al.,}{{Venturi} et~al.}{2017}]{Venturi2017}
{Venturi} T.,  et~al., 2017, \mn@doi [\aap] {10.1051/0004-6361/201630014}, \href {https://ui.adsabs.harvard.edu/abs/2017A&A...603A.125V} {603, A125}

\bibitem[\protect\citeauthoryear{{Venturi} et~al.,}{{Venturi} et~al.}{2022}]{Venturi2022}
{Venturi} T.,  et~al., 2022, \mn@doi [\aap] {10.1051/0004-6361/202142048}, \href {https://ui.adsabs.harvard.edu/abs/2022A&A...660A..81V} {660, A81}

\bibitem[\protect\citeauthoryear{{Vollmer}, {Cayatte}, {van Driel}, {Henning}, {Kraan-Korteweg}, {Balkowski}, {Woudt}  \& {Duschl}}{{Vollmer} et~al.}{2001}]{Vollmer2001}
{Vollmer} B.,  {Cayatte} V.,  {van Driel} W.,  {Henning} P.~A.,  {Kraan-Korteweg} R.~C.,  {Balkowski} C.,  {Woudt} P.~A.,   {Duschl} W.~J.,  2001, \mn@doi [\aap] {10.1051/0004-6361:20010116}, \href {https://ui.adsabs.harvard.edu/abs/2001A&A...369..432V} {369, 432}

\bibitem[\protect\citeauthoryear{{Watson} et~al.,}{{Watson} et~al.}{2009}]{Watson2009}
{Watson} M.~G.,  et~al., 2009, \mn@doi [\aap] {10.1051/0004-6361:200810534}, \href {https://ui.adsabs.harvard.edu/abs/2009A&A...493..339W} {493, 339}

\bibitem[\protect\citeauthoryear{{Wayth} et~al.,}{{Wayth} et~al.}{2015}]{Wayth2015}
{Wayth} R.~B.,  et~al., 2015, \mn@doi [\pasa] {10.1017/pasa.2015.26}, \href {https://ui.adsabs.harvard.edu/abs/2015PASA...32...25W} {32, e025}

\bibitem[\protect\citeauthoryear{{Whiteoak}}{{Whiteoak}}{1972}]{Whiteoak1972}
{Whiteoak} J.~B.,  1972, \mn@doi [Australian Journal of Physics] {10.1071/PH720233}, \href {https://ui.adsabs.harvard.edu/abs/1972AuJPh..25..233W} {25, 233}

\bibitem[\protect\citeauthoryear{{Wilber} et~al.,}{{Wilber} et~al.}{2019}]{Wilber2019}
{Wilber} A.,  et~al., 2019, \mn@doi [\aap] {10.1051/0004-6361/201833884}, \href {https://ui.adsabs.harvard.edu/abs/2019A&A...622A..25W} {622, A25}

\bibitem[\protect\citeauthoryear{{Woudt} \& {Kraan-Korteweg}}{{Woudt} \& {Kraan-Korteweg}}{2001}]{Woudt2001}
{Woudt} P.~A.,  {Kraan-Korteweg} R.~C.,  2001, \mn@doi [\aap] {10.1051/0004-6361:20011455}, \href {https://ui.adsabs.harvard.edu/abs/2001A&A...380..441W} {380, 441}

\bibitem[\protect\citeauthoryear{{Woudt}, {Kraan-Korteweg}, {Lucey}, {Fairall}  \& {Moore}}{{Woudt} et~al.}{2008}]{Woudt2008}
{Woudt} P.~A.,  {Kraan-Korteweg} R.~C.,  {Lucey} J.,  {Fairall} A.~P.,   {Moore} S.~A.~W.,  2008, \mn@doi [\mnras] {10.1111/j.1365-2966.2007.12571.x}, \href {https://ui.adsabs.harvard.edu/abs/2008MNRAS.383..445W} {383, 445}

\bibitem[\protect\citeauthoryear{{Yusef-Zadeh}, {Morris}  \& {Chance}}{{Yusef-Zadeh} et~al.}{1984}]{Yusef-Zadeh1984}
{Yusef-Zadeh} F.,  {Morris} M.,   {Chance} D.,  1984, \mn@doi [\nat] {10.1038/310557a0}, \href {https://ui.adsabs.harvard.edu/abs/1984Natur.310..557Y} {310, 557}

\bibitem[\protect\citeauthoryear{{Zamaninasab}, {Savolainen}, {Clausen-Brown}, {Hovatta}, {Lister}, {Krichbaum}, {Kovalev}  \& {Pushkarev}}{{Zamaninasab} et~al.}{2013}]{Zamaninasab2013}
{Zamaninasab} M.,  {Savolainen} T.,  {Clausen-Brown} E.,  {Hovatta} T.,  {Lister} M.~L.,  {Krichbaum} T.~P.,  {Kovalev} Y.~Y.,   {Pushkarev} A.~B.,  2013, \mn@doi [\mnras] {10.1093/mnras/stt1816}, \href {https://ui.adsabs.harvard.edu/abs/2013MNRAS.436.3341Z} {436, 3341}

\bibitem[\protect\citeauthoryear{{de Vaucouleurs}, {de Vaucouleurs}, {Corwin}, {Buta}, {Paturel}  \& {Fouque}}{{de Vaucouleurs} et~al.}{1991}]{RC3}
{de Vaucouleurs} G.,  {de Vaucouleurs} A.,  {Corwin} Herold~G. J.,  {Buta} R.~J.,  {Paturel} G.,   {Fouque} P.,  1991, {Third Reference Catalogue of Bright Galaxies}.
{Springer, New York, NY (USA)}

\bibitem[\protect\citeauthoryear{{de Vos}, {Hatch}, {Merrifield}  \& {Mingo}}{{de Vos} et~al.}{2021}]{deVos2021}
{de Vos} K.,  {Hatch} N.~A.,  {Merrifield} M.~R.,   {Mingo} B.,  2021, \mn@doi [\mnras] {10.1093/mnrasl/slab075}, \href {https://ui.adsabs.harvard.edu/abs/2021MNRAS.506L..55D} {506, L55}

\bibitem[\protect\citeauthoryear{{van Weeren}, {de Gasperin}, {Akamatsu}, {Br{\"u}ggen}, {Feretti}, {Kang}, {Stroe}  \& {Zandanel}}{{van Weeren} et~al.}{2019}]{vanWeeren2019}
{van Weeren} R.~J.,  {de Gasperin} F.,  {Akamatsu} H.,  {Br{\"u}ggen} M.,  {Feretti} L.,  {Kang} H.,  {Stroe} A.,   {Zandanel} F.,  2019, \mn@doi [\ssr] {10.1007/s11214-019-0584-z}, \href {https://ui.adsabs.harvard.edu/abs/2019SSRv..215...16V} {215, 16}

\makeatother
\end{thebibliography}

%%%%%%%%%%%%%%%%%%%%%%%%%%%%%%%%%%%%%%%%%%%%%%%%%%

% Don't change these lines
\bsp	% typesetting comment
\label{lastpage}
\end{document}